\documentclass[twocolumn,times,tighten,twocolappendix]{aastex63}

\newcommand{\civ}{{\sc C iv}}
\newcommand{\ergscm}{$\rm erg~s^{-1}~cm^{-2}$}
\newcommand{\ergscma}{$\rm erg~s^{-1}~cm^{-2}~\AA^{-1}$}
\newcommand{\feii}{Fe {\sc ii}}
\newcommand{\fvar}{$F_{\rm var}$}
\newcommand{\ha}{H$\alpha$}
\newcommand{\hb}{H$\beta$}
\newcommand{\hbi}{H$\beta_{\rm IC}$}
\newcommand{\hbv}{H$\beta_{\rm VBC}$}
\newcommand{\hbt}{H$\beta_{\rm tot}$}
\newcommand{\hg}{H$\gamma$}
\newcommand{\hei}{He {\sc i}}
\newcommand{\heii}{He {\sc ii}}
\newcommand{\kms}{$\rm km~s^{-1}$}
\newcommand{\lya}{Ly$\alpha$}
\newcommand{\mgii}{Mg {\sc ii}}
\newcommand{\oii}{[O {\sc ii}]}
\newcommand{\oiii}{[O {\sc iii}]}
\newcommand{\rmax}{$r_{\rm max}$}
\newcommand{\sline}{$\sigma_{\rm line}$}

\shorttitle{Two BLR components in PG 0026+129}
\shortauthors{Hu et al.}

\begin{document}

\title{Evidence for Two Distinct Broad-Line Regions from Reverberation Mapping
of PG 0026+129}

\correspondingauthor{Jian-Min Wang}
\email{wangjm@ihep.ac.cn}

\author{Chen Hu}
\affil{Key Laboratory for Particle Astrophysics, Institute of High Energy
Physics, Chinese Academy of Sciences, 19B Yuquan Road, Beijing 100049, China}

\author{Sha-Sha Li}
\affil{Key Laboratory for Particle Astrophysics, Institute of High Energy
Physics, Chinese Academy of Sciences, 19B Yuquan Road, Beijing 100049, China}
\affil{School of Astronomy and Space Science, University of Chinese Academy of
Sciences, 19A Yuquan Road, Beijing 100049, China}

\author{Wei-Jian Guo}
\affil{Key Laboratory for Particle Astrophysics, Institute of High Energy
Physics, Chinese Academy of Sciences, 19B Yuquan Road, Beijing 100049, China}
\affil{School of Astronomy and Space Science, University of Chinese Academy of
Sciences, 19A Yuquan Road, Beijing 100049, China}

\author{Sen Yang}
\affil{Key Laboratory for Particle Astrophysics, Institute of High Energy
Physics, Chinese Academy of Sciences, 19B Yuquan Road, Beijing 100049, China}
\affil{School of Astronomy and Space Science, University of Chinese Academy of
Sciences, 19A Yuquan Road, Beijing 100049, China}

\author{Zi-Xu Yang}
\affil{Key Laboratory for Particle Astrophysics, Institute of High Energy
Physics, Chinese Academy of Sciences, 19B Yuquan Road, Beijing 100049, China}
\affil{School of Astronomy and Space Science, University of Chinese Academy of
Sciences, 19A Yuquan Road, Beijing 100049, China}

\author{Dong-Wei Bao}
\affil{Key Laboratory for Particle Astrophysics, Institute of High Energy
Physics, Chinese Academy of Sciences, 19B Yuquan Road, Beijing 100049, China}
\affil{School of Astronomy and Space Science, University of Chinese Academy of
Sciences, 19A Yuquan Road, Beijing 100049, China}

\author{Bo-Wei Jiang}
\affil{Key Laboratory for Particle Astrophysics, Institute of High Energy
Physics, Chinese Academy of Sciences, 19B Yuquan Road, Beijing 100049, China}
\affil{School of Astronomy and Space Science, University of Chinese Academy of
Sciences, 19A Yuquan Road, Beijing 100049, China}

\author{Pu Du}
\affil{Key Laboratory for Particle Astrophysics, Institute of High Energy
Physics, Chinese Academy of Sciences, 19B Yuquan Road, Beijing 100049, China}

\author{Yan-Rong Li}
\affil{Key Laboratory for Particle Astrophysics, Institute of High Energy
Physics, Chinese Academy of Sciences, 19B Yuquan Road, Beijing 100049, China}

\author{Ming Xiao}
\affil{Key Laboratory for Particle Astrophysics, Institute of High Energy
Physics, Chinese Academy of Sciences, 19B Yuquan Road, Beijing 100049, China}

\author{Yu-Yang Songsheng}
\affil{Key Laboratory for Particle Astrophysics, Institute of High Energy
Physics, Chinese Academy of Sciences, 19B Yuquan Road, Beijing 100049, China}
\affil{School of Astronomy and Space Science, University of Chinese Academy of
Sciences, 19A Yuquan Road, Beijing 100049, China}

\author{Zhe Yu}
\affil{Key Laboratory for Particle Astrophysics, Institute of High Energy
Physics, Chinese Academy of Sciences, 19B Yuquan Road, Beijing 100049, China}
\affil{School of Astronomy and Space Science, University of Chinese Academy of
Sciences, 19A Yuquan Road, Beijing 100049, China}

\author{Jin-Ming Bai}
\affil{Yunnan Observatories, The Chinese Academy of Sciences, Kunming 650011,
China}

\author{Luis C. Ho}
\affil{Kavli Institute for Astronomy and Astrophysics, Peking University,
Beijing 100871, China}
\affil{Department of Astronomy, School of Physics, Peking University, Beijing
100871, China}

\author{Wei-Hao Bian}
\affil{Physics Department, Nanjing Normal University, Nanjing 210097, China}

\author{Michael S. Brotherton}
\affil{Department of Physics and Astronomy, University of Wyoming, Laramie, WY
82071, USA}

\author{Ye-Fei Yuan}
\affil{Department of Astronomy, University of Science and Technology of China,
Hefei 230026, China}

\author{Jes\'us Aceituno}
\affil{Centro Astronomico Hispano Alem\'an, Sierra de los filabres sn, 04550
gergal.  Almer\'ia, Spain}
\affil{Instituto de Astrof\'isica de Andaluc\'ia (CSIC), Glorieta de la
astronom\'ia sn, 18008 Granada, Spain}

\author{Hartmut Winkler}
\affil{Department of Physics, University of Johannesburg, PO Box 524, 2006
Auckland Park, South Africa}

\author{Jian-Min Wang}
\affil{Key Laboratory for Particle Astrophysics, Institute of High Energy
Physics, Chinese Academy of Sciences, 19B Yuquan Road, Beijing 100049, China}
\affil{School of Astronomy and Space Science, University of Chinese Academy of
Sciences, 19A Yuquan Road, Beijing 100049, China}
\affil{National Astronomical Observatories of China, The Chinese Academy of
Sciences, 20A Datun Road, Beijing 100020, China}

\collaboration{20}{(SEAMBH collaboration)}

\begin{abstract}

  We report on the results of a new spectroscopic monitoring campaign of
  the quasar PG 0026+129 at the Calar Alto Observatory 2.2m telescope from
  July 2017 to February 2020. Significant variations in the fluxes of
  the continuum and broad-emission lines, including \hb\ and \heii, were
  observed in the first and third years, and clear time lags between
  them are measured. The broad \hb\ line profile consists of two Gaussian
  components: an intermediate-width \hbi\ with a full width at half-maximum
  (FWHM) of 1964$\pm$18 \kms\ and another very broad \hbv\ with a FWHM
  of 7570$\pm$83 \kms. \hbi\ has long time lags of $\sim$40--60
  days in the rest frame, while \hbv\ shows nearly zero time delay with
  respect to the optical continuum at 5100 \AA.  The velocity-resolved delays
  show consistent results: lags of $\sim$30--50 days at the core of the
  broad \hb\ line and roughly zero lags at the wings. \hbi\ has a redshift of
  $\sim$400 \kms\ which seems to be stable for nearly 30 years by comparing
  with archived spectra, and may originate from an infall. The root mean
  square (rms) spectrum of \hbv\ shows a double-peaked profile with brighter
  blue peak and extended red wing in the first year, which matches the
  signature of a thin disk. Both the double-peaked profile and the
  near-zero lag suggest that \hbv\ comes from a region associated with the
  part of the accretion disc that emits the optical continuum. Adopting
  the FWHM (in the rms spectrum) and the time lag measured for the total \hb\
  line, and a virial factor of 1.5, we obtain a virial mass of
  $2.89_{-0.69}^{+0.60} \times10^7 M_{\odot}$ for the central black hole in
  this quasar.

\end{abstract}


\section{Introduction}
\label{sec-intro}

The broad emission lines are one of the most prominent features of active
galactic nuclei (AGNs) (see \citealt{gaskell09} for a review). They are widely
used for classification (type I/II and the unification model, e.g.,
\citealt{antonucci93}; and narrow-line Seyfert 1 galaxies, NLS1s, e.g.,
\citealt{osterbrock85}), measuring the mass of the central black hole (by both
reverberation mapping and a single-epoch spectrum, see \citealt{peterson14}
for a review), and studying the physics (by the so called Eigenvector 1,
\citealt{boroson92}) and evolution (e.g., \citealt{wang12}) of AGNs. \hb\ is
the most studied broad line in the optical spectra of AGNs. However, the
geometry and kinematics of its emitting region are still far from well
understood. Various complex physical processes and dynamics other than
simple virial motions have been suggested to characterize this region, e.g.,
wind \citep{murry95}, inflow \citep{zhou19} and outflow \citep{czerny11}, and
tidal disruption of clumps from the dusty torus \citep{wang17}.
The observational evidence is often hard to be thoroughly interpreted for
two reasons: the profiles of \hb\ broad lines in different objects are highly
diverse, and the variability behavior of the line in a single object is also
complex and changeable on a time scale of a few years. 

The profiles of the broad \hb\ emission lines are often shifted relative to
the narrow lines and more or less asymmetric \citep[e.g.][]{boroson92}, and
generally cannot be described well by a single simple analytical function
(e.g., a Gaussian or Lorentzian, see \citealt{hu12} for a review), indicating
that multiple physically distinct components may exist in the \hb-emitting
region. The two-component model for AGNs, including an intermediate-width
component and a very broad component,%
\footnote{Note that the terms intermediate-width component and very broad
component have been used in the literature for different emission lines
including not only \hb\ but also \mgii, \civ\ etc.
\citep[e.g.,][]{brotherton94,sulentic99,popovic19}. In this paper, they refer
to the \hb\ emission line only.}
has been proposed in the literature by many authors, as a spectral
decomposition of the \hb\ profile to two Gaussians or a Gaussian plus a
Lorentzian
\citep[e.g.][]{corbin95,brotherton96,popovic04,hu08a,marziani09,kovacevic10}.
However, the emission-line profile emitted from a single physical
component is not necessarily a simple Gaussian or Lorentzian. E.g., a thin
disk is believed to emit an asymmetric double-peaked profile, which has been
observed in many objects \citep{eracleous94,strateva03,storchi-bergmann17}. So
spectral principal component analysis, as an model-independent approach, has
been performed to samples of quasar spectra, and the scenario of two
kinematically distinct \hb\ components is still favored \citep{hu12}.

On the other hand, a scenario of multiple broad-line regions (BLRs) has been
suggested from a theoretical perspective by many authors. E.g.,
\citet{netzer10} found that at least two populations of clouds are required in
their calculations to reproduce the observed line profiles, and simple models
with only one-zone can be ruled out. \citet{wang14} calculated the
self-shadowing of the super-Eddington accreting slim disk, suggested the
possible existence of two BLRs. Numerical simulations of photoionized gas by
\citet{adhikari16} showed the production of intermediate-width emission lines
in high density gas because of the inefficient dust suppression.

More interestingly, multiple \hb\ broad-line components are also expected for
supermassive binary black hole systems, in which the two components relate to
different central black holes and have different velocity shifts
\citep[e.g.][]{boroson09}. Alternatively, for rapidly recoiling black holes
\citep[e.g.][]{eracleous12}, one black hole is ``kicked'' out during the
merger and has a large offset velocity relative to the host galaxy. Thus,
besides searching the evidence of multiple emission-line components by
statistics in large samples, it is more valuable and convincing to identify
such a phenomenon in individual objects. And furthermore, it is important to
distinguish whether the multiple components are emitted from physically
distinct regions of a single black hole, or related to different black holes.

The identification of multiple emission-line components in individual AGNs is
often achieved by performing multi-epoch spectroscopic observations and
recognizing the independent variations of different components. E.g.,
\citet{sulentic00} observed in quasar PG 1416$-$129 that the narrower
``classical'' broad \hb\ component declined dramatically while the very broad
component persisted in two spectra taken ten years apart. For binary
black hole systems, not only the strengths but also the velocity shifts of the
emission-line components are expected to have periodical variabilities. By
decomposing the broad-emission line to several components and studying the
variations in their velocity shifts from long-term spectroscopic monitoring,
several supermassive black hole binary systems have been claimed in the
literature, e.g., NGC 4151 by \citet{bon12}, NGC 5548 by \citet{li16} and
\citet{bon16}.

The existence of multiple BLRs is more convincing if they are not only
kinematically distinct as suggested above, but also geometrically
separated. An explicit result of the virialization of two populations of
clouds with different velocities is that they will rotate at different
distances to the central supermassive black hole. Thus, the reverberation
mapping \citep{blandford82} method would be able to identify multiple
broad-line components by detecting different time lags between
the variations of separated components and that of the ionization continuum.
The time lag represents the distance, because it is the time the ionization
photons travel from the central continuum source to the ionized line-emitting
gas. \citet{bian10} and \citet{zhang13} attempted to decompose the \hb\ broad
line to two components, and then measure the time lag of each by reanalyzing
the spectroscopic monitoring data of \citet{kaspi00} for PG 1700+518 and PG
0052+251, respectively. But the quality of the data sets (mainly the low
sampling cadence) allowed no significant detection of two well separated time
lags.

Similar to the complexity of the line profile discussed above, a single
emission-line component does not need to have a single time lag. It more
probably shows complex structure on the velocity-delay map, which could be
recovered from high quality data (by, e.g., the maximum-entropy method;
\citealt{horne94}). The results will be much more complicated in the case of
two kinematically distinct BLRs with comparable size as simulated in
\citep{wang18}, although the current data quality is not good enough to reveal
such fine structures. The velocity-delay maps, or at least high quality
velocity-resolved delays, have been measured for many objects in several
campaigns
\citep[e.g.][]{denney09,bentz09,grier13b,du16,du18a,xiao18a,derosa18}, showing
diverse kinematic signatures including outflowing, infalling, and virialized
motion. It turns out that the kinematics drawn from the velocity-delay map
does not always agree with that given by profile decomposition. The
high-quality velocity-delay map of NGC 5548 recovered by \citet{xiao18b}
suggests a Keplerian disk, rather than two separated BLRs in the scenario of a
supermassive black hole binary preferred by the profile analysis mentioned
above. Even an intermediate-width long-lag component and a very broad
short-lag component do not have to be emitted from two geometrically separated
BLRs. E.g., in a model of circular Keplerian orbits shown in Figure 10a of
\citet{bentz09}, if the line profile is decomposed to two components, one
representing the line core and another covering the wings, then the core
component will have a longer lag than the wing component. Thus, conclusions
drawn from only profile analysis should be reexamined with caution, by taking
at least velocity-resolved delays into consideration.

Recently, the direct modeling method \citep{pancoast11} has been developed
and applied to roughly a dozen AGNs \citep[e.g.][]{pancoast14,grier17a,li18}
to measure the black hole masses. By establishing a theoretical model of the
motions of the BLR clouds and fitting the yielded line profile variations to
the observed reverberation mapping data, the geometry and kinematics of the
BLR can be constrained. \citet{li18} found that a two-zone model is favored
for Mrk 142, although a complicated one-zone model still can't be ruled out.
With improvements in both the observed data and theoretical modeling (see
\citealt{mangham19} for a discussion), this method would hopefully be able to
confirm the existence of two distinct dynamical components, and distinguish
whether they come from two BLRs of a single black hole or just different black
holes in a binary.

This paper presents strong observational evidence for the existence of
two separated \hb-emitting regions in the quasar PG 0026+129. Both the profile
decomposition and velocity-resolved delays suggest that the very broad
component is emitted from a disk adjacent to the optical continuum source,
while the intermediate-width component probably originates from an infall far
away. Section \ref{sec-obs} briefly describes the observations and data
reduction of our recent spectroscopic monitoring of this object using
the Centro Astron\'omico Hispano-Alem\'an (CAHA) 2.2m telescope. Section
\ref{sec-lc} presents the spectral decomposition and the measurements of light
curves. The analysis to the light curves and the velocity-resolved delays are
then given in Section \ref{sec-result}. The properties and possible origins of
the two broad \hb\ components, along with an estimation of the black hole
mass, are discussed in Section \ref{sec-discu}. Section \ref{sec-sum} gives a
brief summary. 

\section{Observations and Data Reduction}
\label{sec-obs}

PG 0026+129 is a bright radio-quiet quasar with $V$-band magnitude of 15.4 and
redshift $z$=0.1454.%
\footnote{This value is given by the shift of the \oiii\ $\lambda$5007 line in
our spectra, and is slightly larger than that in the NASA/IPAC Extragalactic
Database (NED; \url{http://ned.ipac.caltech.edu/}). See Sections
\ref{sec-fit} for details.}
The full width at half-maximum (FWHM) of its broad \hb\ line given by
\citet{boroson92} is 1860 \kms, allowing it to be classified as an NLS1. With
regards to the other typical spectral features of NLS1, the \feii\ emission is
just moderately strong while \oiii\ lines are far from weak in this object
(see Figure \ref{fig-intgwin} below for an impression). The \hb\ profile
clearly has strong broad wings.

PG 0026+129 has been monitored once by \citet{kaspi00}, and a time lag of
$125_{-36}^{+29}$ days was obtained for its \hb. Only 56 spectroscopic epochs
were observed in $\sim$7.5 years. Data with such a low cadence should be dealt
with caution as the time lag measured could be overestimated by
undersampling, as in the case of PG 2130+099 (see discussions in
\citealt{grier08} and \citealt{hu20}). Moreover, obtaining results beyond
an averaged time lag, e.g., velocity-resolved delays, requires
higher sampling cadence. 

Since May 2017, we started a large reverberation mapping campaign using the
CAHA 2.2m telescope at the Calar Alto Observatory in Spain, which is still
ongoing. This campaign is an expansion of the super-Eddington accreting
massive black hole (SEAMBH) project \citep{du14}, performing long-term and
high-cadence spectroscopic monitoring of PG quasars (PG refers to the
Palomar-Green Survey; \citealt{schmidt83}) with high accretion rates. The
first result of this campaign has been presented in \citet{hu20}, on an
unexpected change of the BLR structure in PG 2130+099 during only two years.
The details of the observations and data reduction of this campaign have been
described in \citet{hu20}, so we only briefly present those relevant to PG
0026+129 below.

PG 0026+129 was observed for 47 epochs between July 2017 and February 2018
(hereafter the observations in 2017), 41 epochs between June 2018 and
February 2019 (hereafter the observations in 2018), and 39 epochs between
August 2019 and February 2020 (hereafter the observations in 2019). For each
epoch, broad-band images and long-slit spectra were taken by the Calar Alto
Faint Object Spectrograph (CAFOS). On the spectrophotometric flux calibration,
we follow the strategy in \citet{kaspi00} and \citet{du14}, that requires
rotating the slit to observe a nearby comparison star with the object
simultaneously. In this case, the comparison star is a G-type (determined from
our spectra) star, located 95$\arcsec$ away from PG 0026+129 with a position
angle of 42$\degr$.

\subsection{Photometry}
\label{sec-phot}

Utilizing the advantage that CAFOS can swiftly switch observing modes between
direct imaging and spectroscopy, broad band images were also taken for two
purposes: 1) confirming that the comparison star is non-varying; 2) comparing
the light curves of the object from photometry and spectroscopy to test the
spectrophotometric flux calibration. For each epoch, three exposures of 60 s
each were taken with a Johnson $V$ filter. Data reduction followed the
standard IRAF%
\footnote{IRAF is distributed by the National Optical Astronomy Observatories,
which are operated by the Association of Universities for Research in
Astronomy, Inc., under cooperative agreement with the National Science
Foundation.}
procedures, and differential instrumental magnitudes of both the object and
the comparison star were obtained relative to the other stars within the
field.

\begin{figure}
  \centering
  \includegraphics[width=0.475\textwidth]{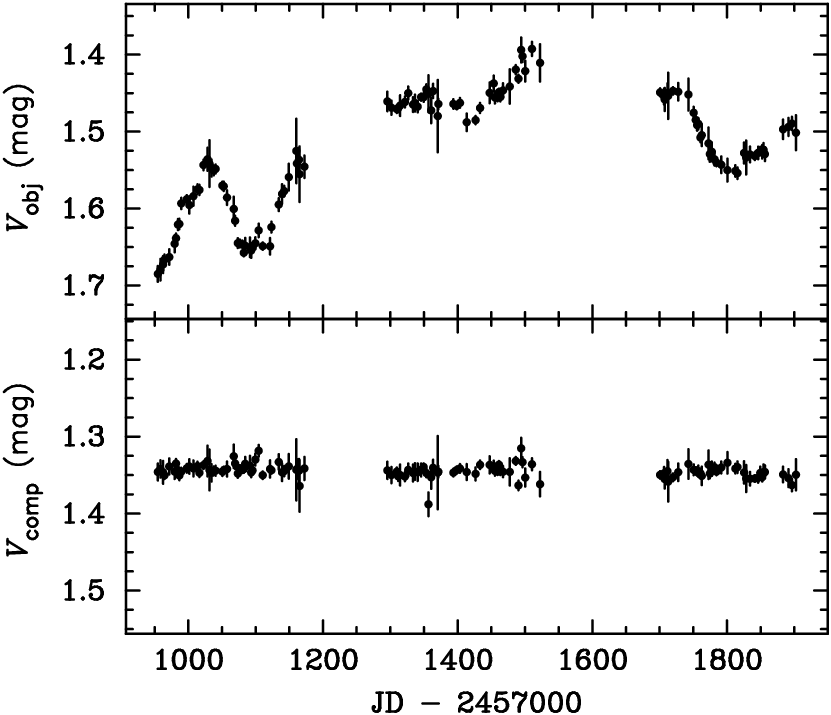}
  \caption{Photometric $V$-band light curves for PG 0026+129 (top) and the
  comparison star (bottom), in units of instrumental magnitudes.
  }
  \label{fig-lcphot}
\end{figure}

Figure \ref{fig-lcphot} shows the $V$-band light curves for PG 0026+129 (top)
and its comparison star (bottom), respectively. The comparison star is rather
stable as the scatter in its magnitudes is only $\lesssim$0.01 mag. The
variability of PG 0026+129 shows significant structure in 2017 and 2019, with
amplitudes of nearly 0.2 and 0.1 mag between the maximum and minimum,
respectively. But in 2018, PG 0026+129 is almost non-varying in the first
$\sim$170 days, and the standard deviation of its magnitudes for the whole
year is only $\sim$0.02 mag.

\subsection{Spectroscopy}

For each epoch, two successive exposures of 1200 s were taken using CAFOS with
Grism G-200 and a long slit with a projected width of 3$\farcs$0. The
spectroscopic images were reduced following the standard procedures using IRAF
(see \citealt{hu20} for details), and then the spectra of both PG 0026+129 and
the comparison star were extracted in a uniform aperture of 10$\farcs$6. The
yielded spectra cover the wavelength range of 4000--8500 \AA\ with a
dispersion of 4.47 \AA\ pixel$^{-1}$. Because the slit is broader than the
seeing most of the time, the actual spectral resolution is better than that
given by the line width of the wavelength-calibration lamp spectra, and
varies in different exposures depending on the seeing. By comparing the widths
of \oiii\ $\lambda$5007 emission lines in our mean spectra with those given by
previous high-spectral-resolution observations for several objects in our
campaign, we estimated a FWHM of 1000 \kms\ as the average instrument
broadening \citep{hu20}. The signal-to-noise ratio (S/N) of PG 0026+129
typically reaches $\sim$80 per pixel at the continuum around the rest-frame
5100 \AA\ for a single exposure. 

The flux calibration was done by using the comparison star as a
spectrophotometric standard. The details of generating the fiducial spectrum
of the comparison star, fitting the sensitivity function, and performing the
calibration to the object spectra were described in \citet{hu20}. The accuracy
of the flux calibration by this technique has been proved to be better than
$\sim$3\% \citep[e.g.][]{kaspi00,hu20}, and the issue of apparent flux
variations of the host galaxy \citep{hu15} can be ignored in the case of PG
0026+129 due to its weak host contribution.

The spectra of six epochs are removed from the following light curve
measurements, because of S/N lower than 20 (per pixel around the
rest-frame 5100 \AA), difference between the fluxes of the two exposures
larger than 3\%, or abnormal spectral slope, all of which were due to bad
weather conditions. Thus, the spectroscopic light curves below contain 46
epochs in 2017, 39 epochs in 2018, and 36 epochs in 2019, respectively.

\section{Light Curve Measurements}
\label{sec-lc}

Two methods are often used in reverberation mapping studies for light curve
measurements: integration and spectral fitting. Integration is traditional,
and widely used in most campaigns
\citep[e.g.,][]{kaspi00,peterson04,bentz09,fausnaugh17,du18a,du18b} for its
simplicity and robustness under normal situations. By subtracting the
continuum as a straight line defined by two windows, the emission-line flux is
measured by a simple integration in a window. This method is suitable for
strong, single emission lines such as \ha\ and \hb, as long as the continuum
can be approximated by a straight line define by the two windows. The
spectral fitting technique is relatively new and adopted in fewer
reverberation mapping campaigns \citep[e.g.,][]{barth13,hu15,barth15}. The
fluxes of emission lines are obtained from fitting the spectra in a wide
wavelength range by including as many spectral components as necessary. This
method is useful especially in two situations: 1) for highly blended lines,
e.g., \feii\ and \heii, spectral fitting is necessary to decompose them from
contaminations \citep{bian10,barth13,hu15}; 2) for objects with strong host
contribution, the continuum deviates considerably from a simple straight line
\citep{hu15,hu16}. However, spectral fitting is less robust than integration
if the quality of the spectra is not high enough for reliably determining each
spectral component, especially the host starlight. In \citet{hu20}, both
methods were used: integration for \hb\ and \hei, and spectral fitting for
\feii\ and \heii.

In this section, we firstly investigate the uncertainty in measuring the \hb\
light curve by the integration method for PG 0026+129, caused by the
contamination to the continuum by the broad \heii\ line. Then, we describe the
spectral fitting method which is preferred in this case due to its better
continuum subtraction. Moreover, we obtain the light curves of the two \hb\
components decomposed by spectral fitting, both of which seem to have physical
meaning.

\subsection{Integration}
\label{sec-intg}

\begin{figure}
  \centering
  \includegraphics[angle=-90,width=0.475\textwidth]{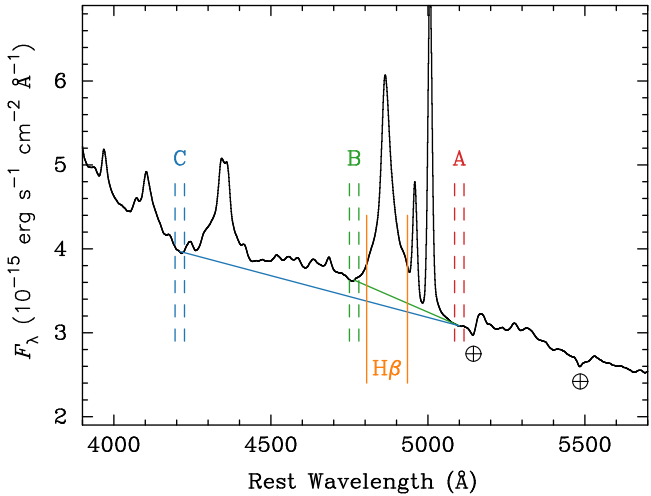}
  \caption{Integration windows for continuum (A, B, and C, between the
  corresponding pairs of dashed vertical lines) and \hb\ line (between the
  solid orange vertical lines). The green and blue solid lines show the
  continua defined by windows A \& B and A \& C, respectively. The spectrum is
  the mean spectrum, and two telluric absorptions are marked by
  $\oplus$.
  }
  \label{fig-intgwin}
\end{figure}

For integrating the flux of \hb, two continuum windows are used to define a
local continuum. The best choice for the red-ward continuum window is at 5100
\AA\ in the rest frame, in which only very weak \feii\ emission is likely
above the continuum (window A in Figure \ref{fig-intgwin}, 5085--5115 \AA).
The blue-ward window is usually set just adjacent to the blue wing of \hb, at
the local minimum between \hb\ and \heii\ (window B, 4750--4780 \AA). However,
in this case, the continuum defined by windows A and B is apparently too steep
(the green solid line), because of the contamination of broad \heii\ emission
in window B (see the cyan Gaussian in Figure \ref{fig-spec}). A better choice
could be the local minimum blueward of \hg\ line (window C, 4195--4225
\AA). The continuum defined by this window, which is much further away from
\hb, looks more reasonable (the blue solid line in Figure \ref{fig-intgwin}).

\begin{figure}
  \centering
  \includegraphics[width=0.475\textwidth]{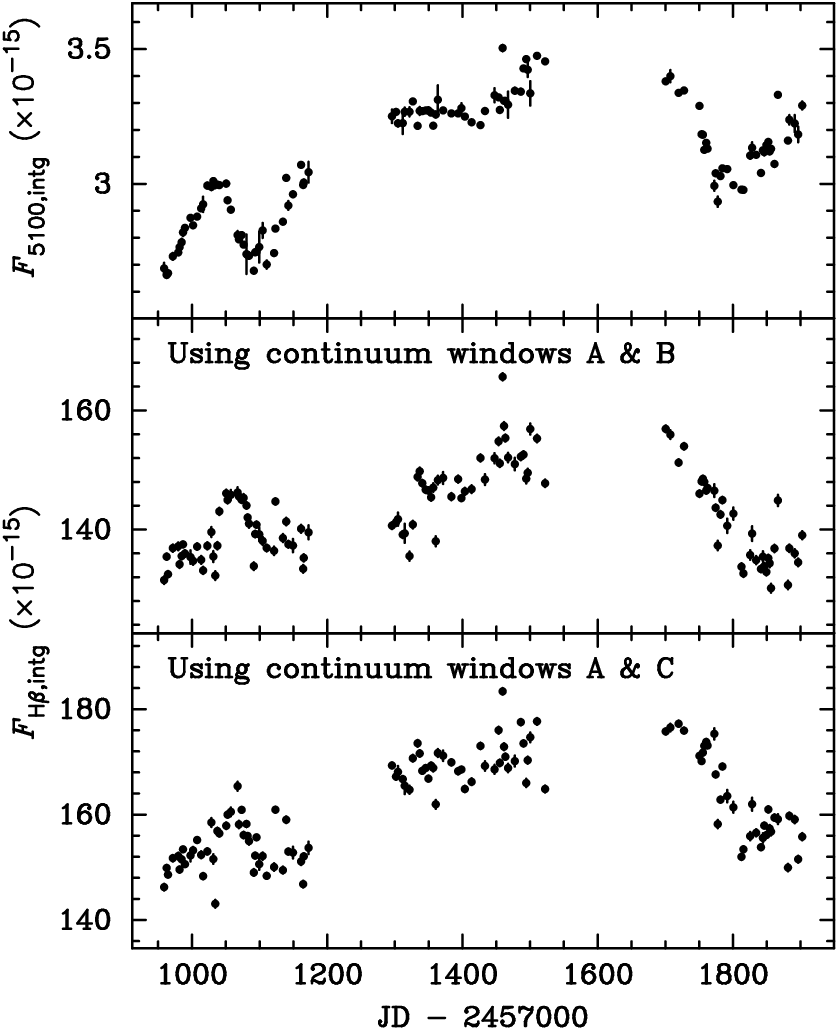}
  \caption{
  Light curves obtained by integration. Top: continuum at rest-frame 5100 \AA.
  Middle and bottom: \hb\ above the continua defined by different choices of
  blue-ward windows. Note the apparent difference between the two \hb\ light
  curves, indicating the uncertainty in integration method. 
  }
  \label{fig-lcintg}
\end{figure}

The top panel of Figure \ref{fig-lcintg} shows the light curve of the
continuum at rest-frame 5100 \AA\ integrated in window A. It agrees well
with that given by $V$-band photometry shown in the top panel of Figure
\ref{fig-lcphot}. The other two panels of Figure \ref{fig-lcintg} show
the \hb\ light curves measured using different choices of the blue-ward
continuum window: by window B in the middle, and C in the bottom. The fluxes
of \hb\ are smaller in the middle panel, because more continuum fluxes are
subtracted as the result of the contamination of \heii\ in window B. Moreover,
the variability of \heii, which is strong as shown in Figure \ref{fig-ccf17},
makes the underestimations of \hb\ fluxes vary in different epochs,
introducing artificial structures in the light curve. This effect is more
severe in 2018, when the variability amplitudes of both continuum and \hb\ are
small, creating the false illusion of increasing of \hb\ flux shown in the
middle panel.

Although window C avoids the contamination of \heii\ emission, its large
wavelength separation from \hb\ increases the systematic error introduced by
the uncertainty in the spectral shape calibration. Thus, the method of spectral
fitting is favored for light curve measurements in this work.

\subsection{Spectral Fitting}
\label{sec-fit}

Our spectral fitting follows that in \citet{hu20}, except that the host
starlight is not included here. In our spectra, there is no recognizable
stellar absorption feature in even the mean spectrum (see Figure
\ref{fig-intgwin}, note that the two broad absorption features around
$\sim$5145 and 5485 \AA\ are telluric absorptions). Also, the continuum shape
in the optical band can be fitted well without adding a starlight component.
Thus, we neglect the host in our spectral fitting.

Before fitting, the Galactic extinction correction was performed with an
extinction law assuming $R_V$=3.1 (\citealt{cardelli89} and
\citealt{odonnell94}) and a $V$-band extinction of 0.195 mag from the NED
determined by \citet{schlafly11}. Then the spectra were deredshifted with a
value of 0.1454, determined by the \oiii\ $\lambda$5007 line in our mean
spectrum.

\begin{figure}
  \centering
  \includegraphics[angle=-90,width=0.475\textwidth]{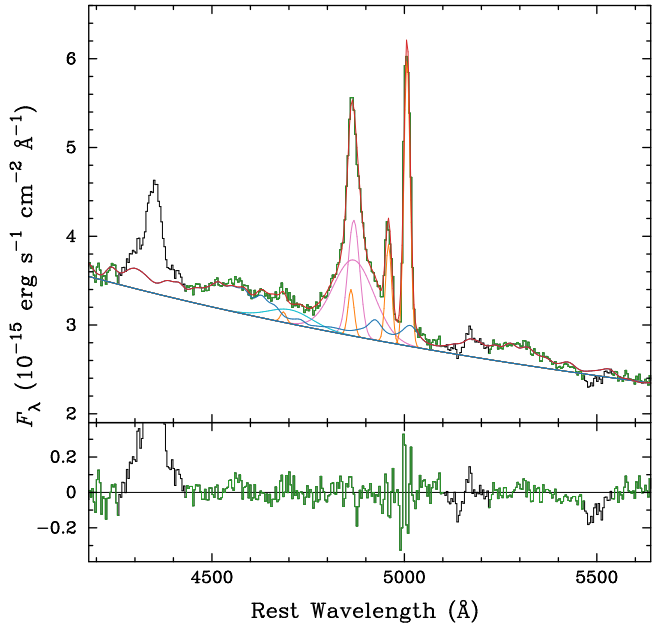}
  \caption{
  Sample fit of a single-epoch spectrum. In the top panel, the observed
  spectrum is plotted in green and black for the pixels included and excluded
  in the fitting, respectively. The best-fit model (red) is the sum of
  the following components: the power-law continuum and the \feii\
  pseudo-continuum (blue), intermediate-width and very broad \hb\
  components (magenta), broad \heii\ line (cyan), and the narrow emission
  lines (orange, including \oiii, \hb, and \heii). The bottom panel shows the
  residuals. Two narrow bands around $\sim$5145 and 5485 \AA\ are excluded in
  the fitting for telluric absorptions.
  }
  \label{fig-spec}
\end{figure}

The following spectral components are included in the fitting, as shown in 
Figure \ref{fig-spec} for a single-epoch spectrum: 1) the AGN continuum
modeled as a simple power law, 2) the \feii\ pseudo-continuum generated by
convolving a Gaussian with the \citet{boroson92} template, 3) the broad \hb\
line as two Gaussians, one for the intermediate-width component (\hbi) and
another for the very-broad component (\hbv), 4) the broad \heii\ as a single
Gaussian, 5) narrow-emission lines including \oiii\ $\lambda\lambda$4959,
5007, \heii\ $\lambda$4686, and \hb, modeled by a set of Gaussians with the
same velocity width and shift.  Because of the blending between the \feii\
emission, broad \heii\ line, and the \hb\ wings, and also the degeneracy
between the two broad \hb\ components, we first fit the mean spectrum with
all the parameters free to vary, and then fit the single-epoch spectra by
fixing the velocity widths and shifts of the broad \heii, \hbi, and \hbv\ to
the values given by the best fit to the mean spectrum. Also, the relative
density ratios between the narrow-emission lines are fixed to the values in
the best-fit model of the mean spectrum. Among the 20 parameters in total,
only 11 are free to vary in the fitting to single-epoch spectra. The fitting
is performed between the rest-frame wavelengths 4180 and 5640 \AA, excluding a
window around \hg\ and two narrow windows around $\sim$5145 and 5485 \AA\ for
telluric absorptions.

The best fit to the mean spectrum yields a non-zero flux of a narrow \hb\
component, whose velocity width and shift are forced to be the same as
those of \oiii. The intensity ratio relative to \oiii\ $\lambda$5007 is
0.156, which is a normal value in AGNs \citep[e.g.,][]{veilleux87}.
Because of the low spectral resolution in this campaign (instrument broadening
$\sim$1000 \kms; \citealt{hu20}), the narrow \hb\ component is smeared with
the red-shifted \hbi, making the peak of the entire \hb\ profile $\sim$150
\kms\ red-shifted with respect to \oiii. It is well known that \oiii, as a
high-ionization line, can be blue-shifted with respect to the low-ionization
lines \citep[e.g.][]{boroson05,hu08b}. We checked the spectra of this object
obtained with the Sutherland 1.9 m telescope at the South African Astronomical
Observatory during this campaign, which have higher spectral resolution (see
Appendix \ref{sec-saao} for details). By comparing the velocity shifts of
\oiii, \oii, and the peak of the \hb\ profile, we concluded that the \oiii\
lines of PG 0026+129 are not blue-shifted with respect to the low-ionization
lines. Thus, forcing the narrow \hb\ component to have the same profile as
\oiii\ in our fitting is appropriate, and so does using \oiii\ to define the
systematic redshift when no host absorption feature is available.

\begin{deluxetable*}{lr@{~$\pm$~}lr@{~$\pm$~}lr@{~$\pm$~}lr@{~$\pm$~}lcr@{}l}
  \tablewidth{0pt}
  \tablecaption{Measurements for Broad-Emission Lines and Components
  \label{tab-results}}
  \tablehead{
  \colhead{Line} & \multicolumn{2}{c}{Flux} & \multicolumn{2}{c}{FWHM} &
  \multicolumn{2}{c}{Shift} & \multicolumn{2}{c}{\fvar} & \colhead{\rmax} &
  \multicolumn{2}{c}{Lag}
  \\
  \colhead{} & \multicolumn{2}{c}{($\times10^{-15}$ \ergscm)} & 
  \multicolumn{2}{c}{(\kms)} & \multicolumn{2}{c}{(\kms)} & 
  \multicolumn{2}{c}{(\%)} & \colhead{} & \multicolumn{2}{c}{(days)}
  \\
  \colhead{(1)} & \multicolumn{2}{c}{(2)} & \multicolumn{2}{c}{(3)} &
  \multicolumn{2}{c}{(4)} & \multicolumn{2}{c}{(5)} & \colhead{(6)} &
  \multicolumn{2}{c}{(7)}
  }
  \startdata
  \multicolumn{12}{c}{2017}
  \\ \tableline
  \heii & 36.0 & 8.5 & 8445 & 162 &
  1173 & 62 & 21.3 & 2.7 & 0.46 &
  $-$1.4 & $_{-6.9}^{+4.9}$
  \\
  \hbv & 124.8 & 7.0 & 7570 & 83 &
  415 & 12 & 4.4 & 0.7 & 0.66 &
  $-$1.9 & $_{-5.5}^{+9.3}$
  \\
  \hbi &  45.8 & 3.2 & 1964 & 18 &
  449 & 3  & 6.7 & 0.7 & 0.81 &
  43.4 & $_{-1.5}^{+4.1}$
  \\
  \hbt & 170.6 & 6.1 & 3193 & 141\tablenotemark{a} &
  424 & 1  & 2.5 & 0.5 & 0.61 &
  11.7 & $_{-7.8}^{+7.4}$
  \\ \tableline
  \multicolumn{12}{c}{2019}
  \\ \tableline
  \heii & 40.1 & 5.2 & 8445 & 162 &
  1173 & 62 & 11.8 & 1.6 & 0.62 &
  17.8 & $_{-8.4}^{+12.2}$
  \\
  \hbv & 129.4 & 5.5 & 7570 & 83 &
  415 & 12 & 3.9 & 0.5 & 0.74 &
  $-$1.1 & $_{-2.2}^{+12.9}$
  \\
  \hbi &  51.8 & 6.0 & 1964 & 18 &
  449 & 3  & 11.2 & 1.4 & 0.87 &
  60.0 & $_{-11.0}^{+5.9}$\tablenotemark{b}
  \\
  \hbt & 181.2 & 8.7 & 3094 & 133\tablenotemark{a} &
  424 & 1  & 4.6 & 0.6 & 0.87 &
  27.7 & $_{-6.0}^{+5.0}$
  \enddata
  \tablecomments{
  Measurements for the broad \heii\ line, two broad \hb\ components and the
  total broad \hb\ line in years 2017 and 2019. Column (2) lists the
  mean fluxes, and the errors are the standard deviations. Columns (3) and (4)
  list the FWHMs after instrumental broadening correction and the velocity
  shifts with respect to \oiii\ $\lambda$5007, measured from the mean
  spectrum, except those for \hbt, which are the means of the values
  measured from the individual-night spectra. Columns (5) and (6) give the
  variability amplitudes \fvar, and the peak values \rmax\ of the CCFs. And
  column (7) lists the time lags $\tau$ in the rest frame.
  }
  \tablenotetext{a}{
  The value here is the mean of the FWHMs in individual-night spectra
  calculated from the best-fit models, while the FWHMs listed in Table
  \ref{tab-hb} column (2) are those measured directly from the broad-\hb-only
  mean and rms spectra.
  }
  \tablenotetext{b}{
  The time lag of \hbi\ in 2019 could be underestimated here. See the text for
  a discussion.
  }
\end{deluxetable*}

Columns (3) and (4) of Table \ref{tab-results} list the FWHMs and velocity
shifts of the broad-emission lines and components from the best fit to the
mean spectrum.%
\footnote{
Except those for \hbt. They are the mean FWHMs of the individual-night spectra
measured from the best-fit models, in each year. The errors are the standard
deviations.
}
The errors are estimated as the standard deviations of the values given by the
best fits to Monte Carlo realizations (by bootstrap sample selection) of the
mean spectrum. The listed values of FWHMs have been corrected for the
instrumental broadening, and those of velocity shifts are with respect to
\oiii\ $\lambda$5007. Note that \hbv\ is $\sim$3.9 times broader than \hbi.
This width ratio is much higher than the average value of 2.5 in
\citet{hu08a}. The width of \hbv\ approximates that of the broad \heii, while
\hbi\ and \feii\ (FWHM = 1957$\pm$35 \kms, shift = 243$\pm$8 \kms)
have similar widths.

The FWHMs and velocity shifts listed in Table \ref{tab-results} for \heii\ and
two broad \hb\ components are identical for different years because we set
them to the values measured in the mean spectrum for the entire data set. We
tried measuring the mean spectrum of each single year and setting annual
averages separately, but the time lags measured in each year have no
statistically significant change. However, the relative fluxes of these
components between years change, showing some long-term trends. Such trends
are not seen in the results here, which could therefore be caused by the
varying degeneracy in spectral decomposition if different values are used for
different years.

\begin{figure}
  \centering
  \includegraphics[width=0.475\textwidth]{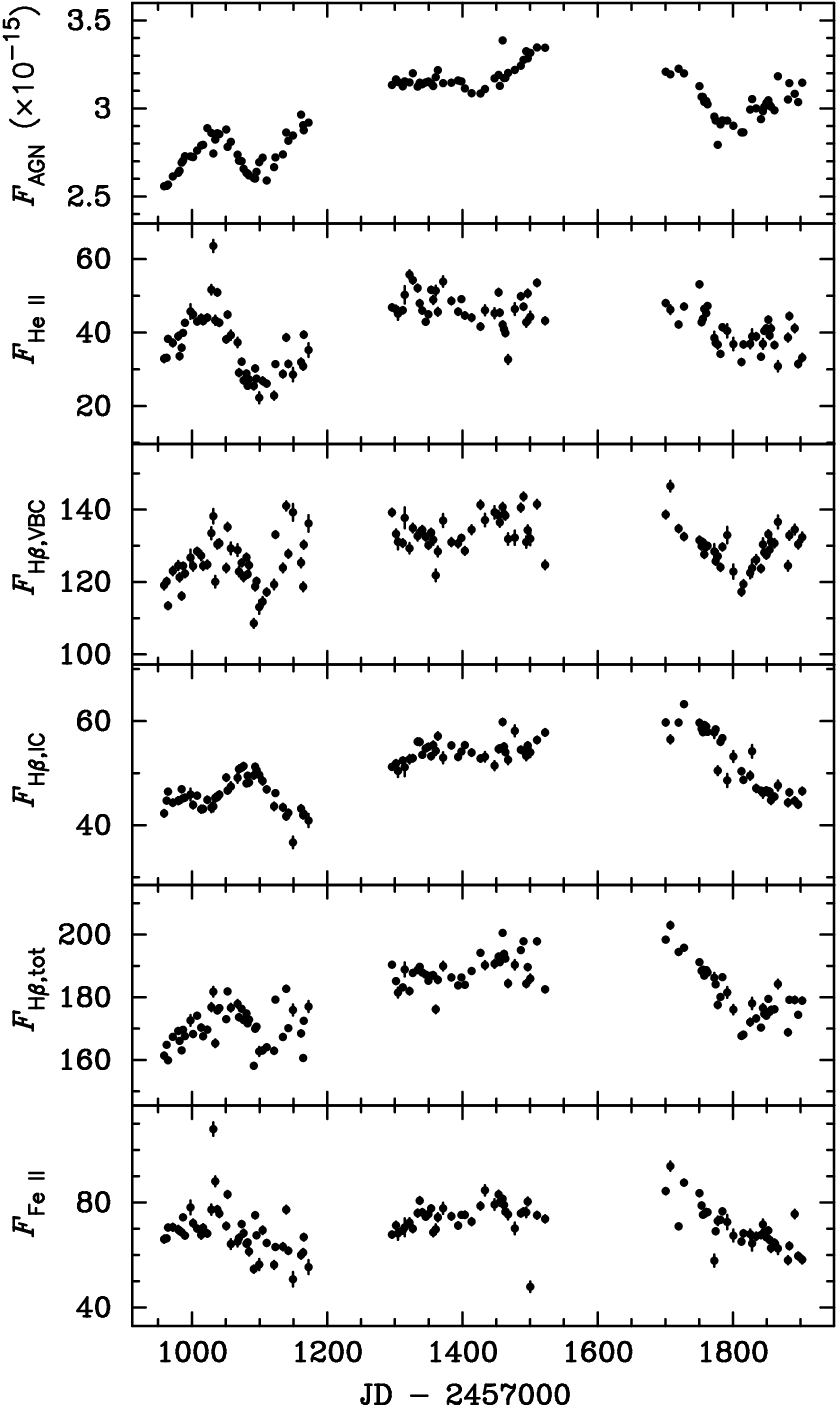}
  \caption{
  Light curves obtained by spectral fitting. From top to bottom: the AGN
  continuum at rest-frame 5100 \AA, the broad \heii, the very-broad \hb\
  component, the intermediate-width \hb\ component, the total broad \hb, and
  the \feii\ emission.
  }
  \label{fig-lcfit}
\end{figure}

\begin{deluxetable*}{cr@{~$\pm$~}lr@{~$\pm$~}lr@{~$\pm$~}lr@{~$\pm$~}lr@{~$\pm$~}lr@{~$\pm$~}l}
  \tablewidth{0pt}
  \tablecaption{Light curves of the 5100 \AA\ continuum and emission lines
  \label{tab-lc}}
  \tablehead{
  \colhead{JD$-$2457000} &
  \multicolumn{2}{c}{$F_{\rm 5100}$} & 
  \multicolumn{2}{c}{$F_{\rm He~\textsc{ii}}$} & 
  \multicolumn{2}{c}{$F_{\rm H\beta,VBC}$} & 
  \multicolumn{2}{c}{$F_{\rm H\beta,IC}$} & 
  \multicolumn{2}{c}{$F_{\rm H\beta,tot}$} & 
  \multicolumn{2}{c}{$F_{\rm Fe~\textsc{ii}}$}
  \\
  \colhead{(1)} & 
  \multicolumn{2}{c}{(2)} & \multicolumn{2}{c}{(3)} &
  \multicolumn{2}{c}{(4)} & \multicolumn{2}{c}{(5)} &
  \multicolumn{2}{c}{(6)} & \multicolumn{2}{c}{(7)} 
  }
  \startdata
  958.663 & 2.559 & 0.004 & 32.92 & 0.97 & 119.1 & 1.2  & 
            42.27 & 0.67  & 161.4 & 0.9  & 66.05 & 1.39 \\
  962.626 & 2.560 & 0.003 & 33.16 & 0.76 & 120.1 & 1.0  & 
            44.72 & 0.55  & 164.9 & 0.8  & 66.36 & 1.10 \\
  964.641 & 2.568 & 0.003 & 38.30 & 0.76 & 113.4 & 1.0  & 
            46.46 & 0.53  & 159.9 & 0.7  & 70.48 & 1.09 \\
  971.608 & 2.614 & 0.004 & 37.17 & 0.94 & 123.1 & 1.2  & 
            44.31 & 0.61  & 167.4 & 0.9  & 70.59 & 1.34 \\
  979.631 & 2.634 & 0.004 & 39.01 & 0.98 & 124.5 & 1.2  & 
            44.71 & 0.66  & 169.3 & 0.9  & 69.70 & 1.42 \\
  \multicolumn{13}{c}{$\vdots$}
  \enddata
  \tablecomments{
  The 5100\AA\ continuum flux is in units of $10^{-15}$ \ergscma, and
  emission-line fluxes are in units of $10^{-15}$ \ergscm.
  (This table is available in its entirety 
  in a machine-readable form on line.)
  }
\end{deluxetable*}

The light curves are generated directly from the fluxes of the decomposed
spectral components given by the best fits to the single-epoch spectra.
Figure \ref{fig-lcfit} shows, from top to bottom, the light curves of the
AGN continuum, the broad \heii, \hbv, \hbi, the total broad \hb\ (\hbt), and
the \feii\ emission. The flux of \hbt\ is just the sum of the fluxes of \hbv\
and \hbi. Table \ref{tab-lc} presents the data of all these light curves
(only the first five epochs are included here as an example; the
machine-readable table in its entirety are available on line).

The mean fluxes of these emission lines and components are listed in column
(2) of Table \ref{tab-results}, and the standard deviations are given as the
errors. The flux ratio of \hbv\ to \hbi\ is $\gtrsim$2.5, thus \hbt\ is
dominated by the former. Note that the error bars plotted in the light curves
are those given directly by the fitting (the statistical errors), and not
adequate to interpret the scatter in the fluxes of successive epochs. Thus, a
systematic error is estimated for each light curve as in \citet{hu15,hu20},
and has been added in quadrature for the time series analysis below.

\section{Time Series Analysis}
\label{sec-result}

We perform time series analysis on the light curves in each single year
separately, to avoid potential influence of the unobservable gaps. On the
other hand, the behavior of emission-line reverberation in different
observing years have been found to be able to change significantly in other
objects (e.g., PG 2130+099; \citealt{hu20}), and are therefore worth exploring
here. For comparison, we present the results of time series analysis performed
on the combined light curves for the entire three years, in Appendix
\ref{sec-3y}. In general, the time lags are consistent with those obtained in
individual years.

As shown in Figure \ref{fig-lcfit}, the variability amplitudes in fluxes of
both AGN continuum and emission lines in 2018 are too small to yield reliable
time lag measurements. Hence, we present only the results for 2017 and 2019
hereafter. In addition, no reliable time lag is obtained for the \feii\
emission in any year, possibly due to the relative larger scatter in its light
curve and potential longer time lag than other lines.

\subsection{Variability Amplitudes}

The quantity \fvar\ and its uncertainty defined by \citet{rodriguez97} and
\citet{edelson02} are calculated to represent the intrinsic variability
amplitude over the errors (including both the statistical and systematic
errors). Column (5) of Table \ref{tab-results} lists the results for the broad
emission lines and components. For comparison, the \fvar\ of the AGN continuum
is 3.8$\pm$0.4\% and 3.3$\pm$0.4\% in 2017 and 2019, respectively. The much
larger \fvar\ of \heii\ compared to those of the continuum and other lines are
commonly seen in previous campaigns \citep[e.g.,][]{barth15,hu20}. Note that
the \fvar\ of \hbt\ is smaller than those of \hbv\ and \hbi\ separately
in 2017, because the variations of the two components are not
synchronous. Also note that the variability amplitude of \hbi\ is much
larger in 2019 than it was in 2017, while \hbv\ (and also the continuum) shows
slightly smaller variability amplitude in 2019, causing the significant change
in the rms spectra of the two years (see Figures \ref{fig-vrd17} and
\ref{fig-vrd19} below).

\subsection{Reverberation lags}
\label{sec-lag}

\begin{figure}
  \centering
  \includegraphics[width=0.475\textwidth]{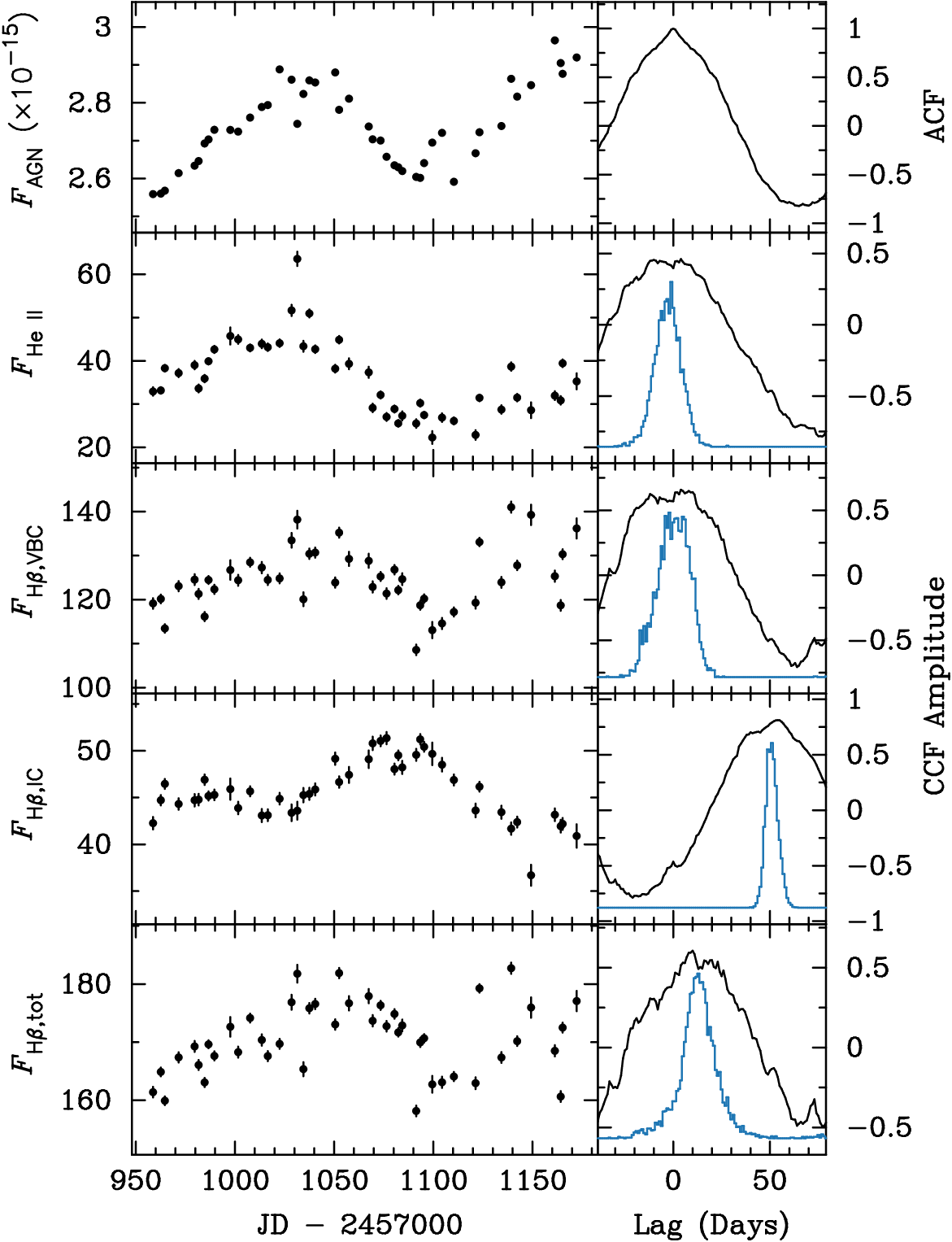}
  \caption{
  Left column, from top to bottom: light curves in 2017 of the AGN continuum
  at 5100 \AA, the broad \heii, the very-broad \hb\ component, the
  intermediate-width \hb\ component, and the total broad \hb\ line from
  spectral fitting. The units for the fluxes of the AGN continuum and emission
  lines are $\times 10^{-15}$ \ergscma\ and $\times 10^{-15}$ \ergscm,
  respectively. Right column: the autocorrelation function of the AGN
  continuum and the cross-correlation functions for the emission lines in the
  left column with respect to the continuum. The blue histograms are the
  corresponding cross-correlation centroid distributions.
  }
  \label{fig-ccf17}
\end{figure}

\begin{figure}
  \centering
  \includegraphics[width=0.475\textwidth]{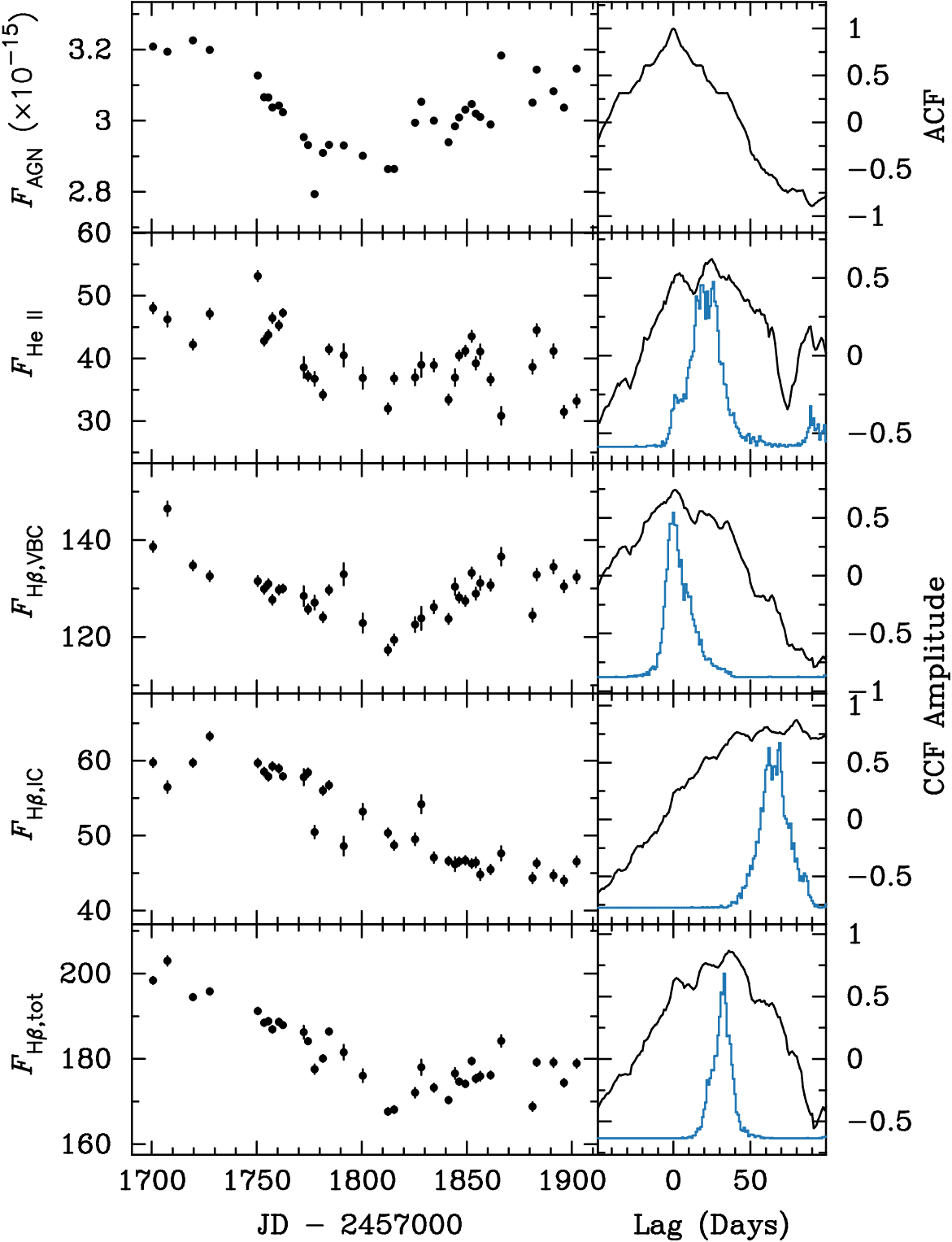}
  \caption{
  Light curves and CCF analysis results in 2019. The units and notations are
  as the same as those in Figure \ref{fig-ccf17}. Note that the time lag of
  \hbi\ could be underestimated here. See the text for a discussion.
  }
  \label{fig-ccf19}
\end{figure}

The reverberation lags between the variations of the emission lines and the
continuum are measured using the standard interpolation cross-correlation
function (CCF) method \citep{gaskell86,gaskell87,white94}.  The value of the
time lag is defined by the centroid of the CCF above the 80\% level of the
peak value (\rmax) following \citet{koratkar91} and \citet{peterson04}. The
uncertainty is in turn estimated by the 15.87\% and 84.13\% quantiles of the
cross-correlation centroid distribution (CCCD) yielded from Monte Carlo
realizations generated by random subset selection/flux randomization
\citep{maoz89,peterson98}. The right columns of Figures \ref{fig-ccf17}
and \ref{fig-ccf19} show the autocorrelation function (ACF) of the AGN
continuum (top panel), the CCFs (in black) and CCCDs (in blue) for the
emission lines and components with respect to the AGN continuum (other
panels), for the years 2017 and 2019 respectively. The \rmax\ and time lags
($\tau$) in the rest frame are listed in columns (6) and (7) of Table
\ref{tab-results}.

In 2017, both \heii\ and \hbv\ have negative values of measured time
lags, with respect to the continuum at 5100\AA. The AGN continuum in this
optical band has been observed to lag behind the ultraviolet (UV) continuum,
which ionizes the line-emitting gas \citep[e.g.,][]{edelson19}. It is expected
if the optical photons are emitted at larger radius on the accretion disc than
the UV photons \citep{cackett07}, and also the contribution of the diffuse
continuum emission from BLR is significant
\citep{korista01,lawther18,chelouche19,korista19,netzer20}. Considering the
median sampling cadence of $\sim$4 days in this campaign and the uncertainties
given by the CCCDs, the time lags of \heii\ and \hbv\ are broadly consistent
with zero, which means that their emitting-region sizes are comparable to the
size of the part of accretion disc that emits the optical continuum. The
negative or nearly zero lags of \heii\ have been reported for many objects in
previous reverberation mapping campaigns \citep[e.g.,][]{barth13}. But such a
short lag of an \hb\ component, containing $\sim$3/4 of the total fluxes, is
totally surprising for such a luminous quasar (see Section \ref{sec-hbi} below
for an estimation of the lag by the BLR radius--luminosity relation).

In 2019, \hbv\ also has a slightly negative time lag consistent with zero as
in 2017. \heii\ shows a rather large time lag of $17.8_{-8.4}^{+12.2}$ days in
2019. However, the light curve of \heii\ shows a smaller variability amplitude
but larger scattering than that in 2017, especially during the second half of
the year. Thus, this change in the time lag of \heii\ between the two years
has to be treated with caution, and this finding should be checked through
future observations.

\hbi\ shows significant lags of $43.4_{-1.5}^{+4.1}$ and $60.0_{-11.0}^{+5.9}$
days (in the rest frame) in 2017 and 2019, respectively. If both \hbv\ and
\hbi\ are virialized, the lag of \hbv\ can be estimated as $\tau$(\hbv)
$\approx$ $\tau$(\hbi) $\times$ [FWHM(\hbi)$/$FWHM(\hbv)]$^2$ $\approx$ 2.9
and 4.0 days in the two years, respectively. These values are consistent with
our measurements of roughly zero, counting the potential time lag between the
ionizing UV photons and the optical photons we observed. The measured time lag
of \hbt\ in 2019 is more than two times as long as that in 2017
($27.7_{-6.0}^{+5.0}$ and $11.7_{-7.8}^{+7.4}$ days, respectively). But both
are roughly equal to the varying-flux-weighted ($F\times F_{\rm var}$) average
of the lags of \hbv\ and \hbi\ in each year. Note that in 2017, the \rmax\ of
\hbt\ is lower than those of both \hbv\ and \hbi, indicating that such a
decomposition in the dynamics of the BLR clouds is also valid in geometry.

Considering the little more than a steady decline of \hbi\ light curve and the
relatively short duration of the $\sim$200 days monitoring in 2019, the
measured lag of $\sim$60 days for \hbi\ can be underestimated. In addition, if
this decline is not just the response to the dimming of the continuum in the
first half of this season but contains a long-term trend in the variability
of only the emission line, the measured time lag becomes much lower than the
current value after subtracting a first-order polynomial to remove this trend
(detrending; \citealt{welsh99}). However, on a longer time scale, the light
curves of the continuum and \hb\ components of the entire three-year data set
do not show different trends (Appendix \ref{sec-3y}). Thus, we prefer to
interpret the 2019 decline in the \hb\ flux as due to reverberation of the
varying continuum, and we choose not to apply detrending.

The large differences in both the velocities (as FWHMs) and the distances to
the central continuum source (as lags) of the two broad \hb\ components,
suggest that they are emitted from two separated regions. As mentioned in
Section \ref{sec-intro}, direct modeling and velocity-resolved delays could
be more convincing in identifying distinct emission-line components. In the
next section, we show the results of velocity-resolved delays, but a direct
modeling study is beyond of the scope of this paper.

\subsection{Velocity-Resolved Delays}
\label{sec-vrd}

\begin{figure}
  \centering
  \includegraphics[width=0.45\textwidth]{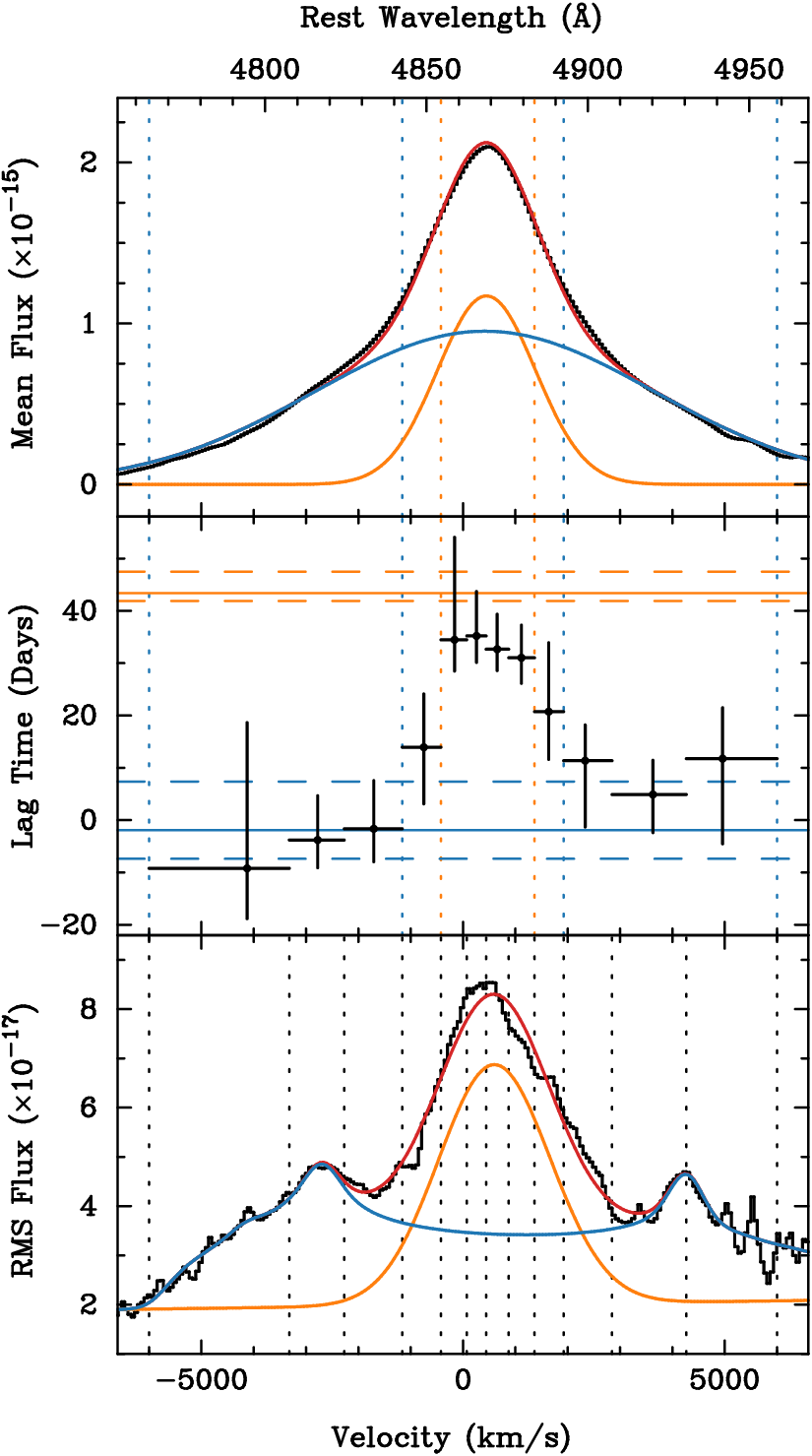}
  \caption{
  Top: the broad-\hb-only mean spectrum (black) and the best-fit model (red)
  consists of \hbv\ (blue) and \hbi\ (orange). Middle: the velocity-resolved
  delays in the rest frame (dots with error bars). The horizontal solid lines
  mark the time lags of \hbv\ (blue) and \hbi\ (orange), and the associated
  dashed lines mark the one sigma errors. In the top and middle panels, the
  dotted blue and orange vertical lines divide the velocity bins into three
  groups: \hbv\ only, \hbi\ dominated, and mixed. Bottom: the broad-\hb-only
  rms spectrum (black) and the best-fit model (red) consists of a Gaussian
  (orange) plus a disk profile (blue). The vertical dotted lines mark the
  boundaries of the bins of equal rms fluxes for measuring the
  velocity-resolved delays. Note that the measurements in the narrow velocity
  bins around the core are not independent due to the instrument broadening.
  }
  \label{fig-vrd17}
\end{figure}

\begin{figure}
  \centering
  \includegraphics[width=0.45\textwidth]{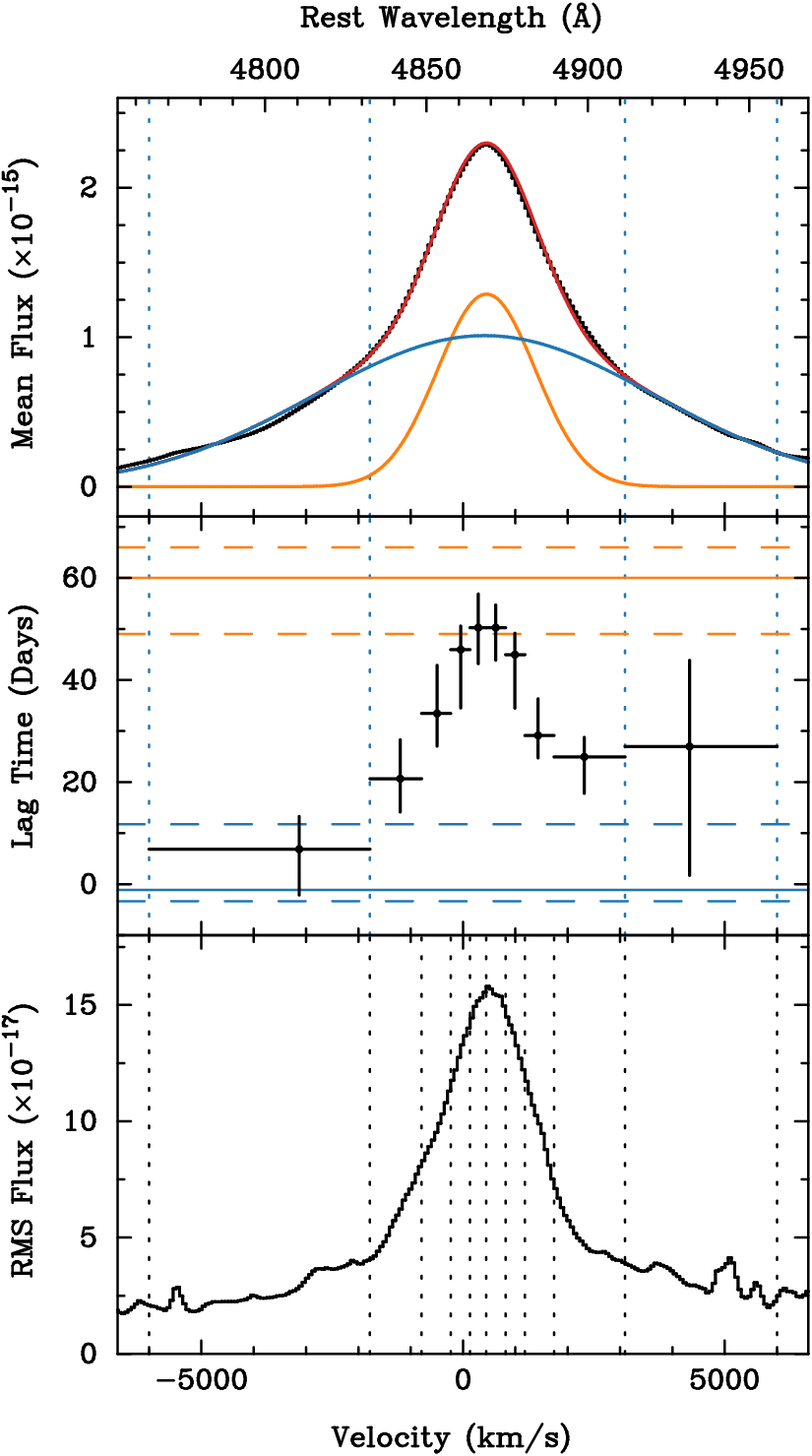}
  \caption{
  The broad-\hb-only mean spectrum (top), the velocity-resolved delays
  (middle), and  the broad-\hb-only rms spectrum (bottom) in 2019. The
  notations are as the same as those in Figure \ref{fig-vrd17}. Note that the
  measurements in the narrow velocity bins around the core are not independent
  due to the instrument broadening.
  }
  \label{fig-vrd19}
\end{figure}

As shown in Section \ref{sec-lc}, spectral fitting is better than simple
integration for determining the continuum and decomposing the contaminations
in this case. Thus, for obtaining the velocity-resolved delays of \hb, we
started with a broad-\hb-only spectrum for each epoch after subtracting the
best-fit models of all other spectral components.%
\footnote{We also generated the velocity-resolved light curves by the
traditional integration from the original spectra. The yielded
velocity-resolved delays show similar features as those given by the
broad-\hb-only spectra here: zero lags at the wings and long lags at the core,
although the uncertainties are larger.}
Then, we generated the root mean square (rms) spectrum of the broad-\hb-only
spectra, and divided it into a dozen bins of equal fluxes between $-$6000 to
6000 \kms\ in the velocity space.%
\footnote{The velocity binning around the line core is below the instrumental
resolution of $\sim$1000 \kms, and thus the measurements near the line core
are not independent.}
The bottom panels of Figures \ref{fig-vrd17}
and \ref{fig-vrd19} show the rms spectra (solid black histogram), and the
boundaries of the velocity bins (dotted vertically lines) in years 2017
and 2019, respectively. Finally, light curves were measured by integrating
the fluxes of the broad-\hb-only spectra in each velocity-space bin, and time
lags were obtained from the CCFs with the AGN continuum light curve given by
the spectral fitting (in the left-top panels of Figures \ref{fig-ccf17}
and \ref{fig-ccf19}).

\begin{figure*}
  \centering
  \includegraphics[width=0.95\textwidth]{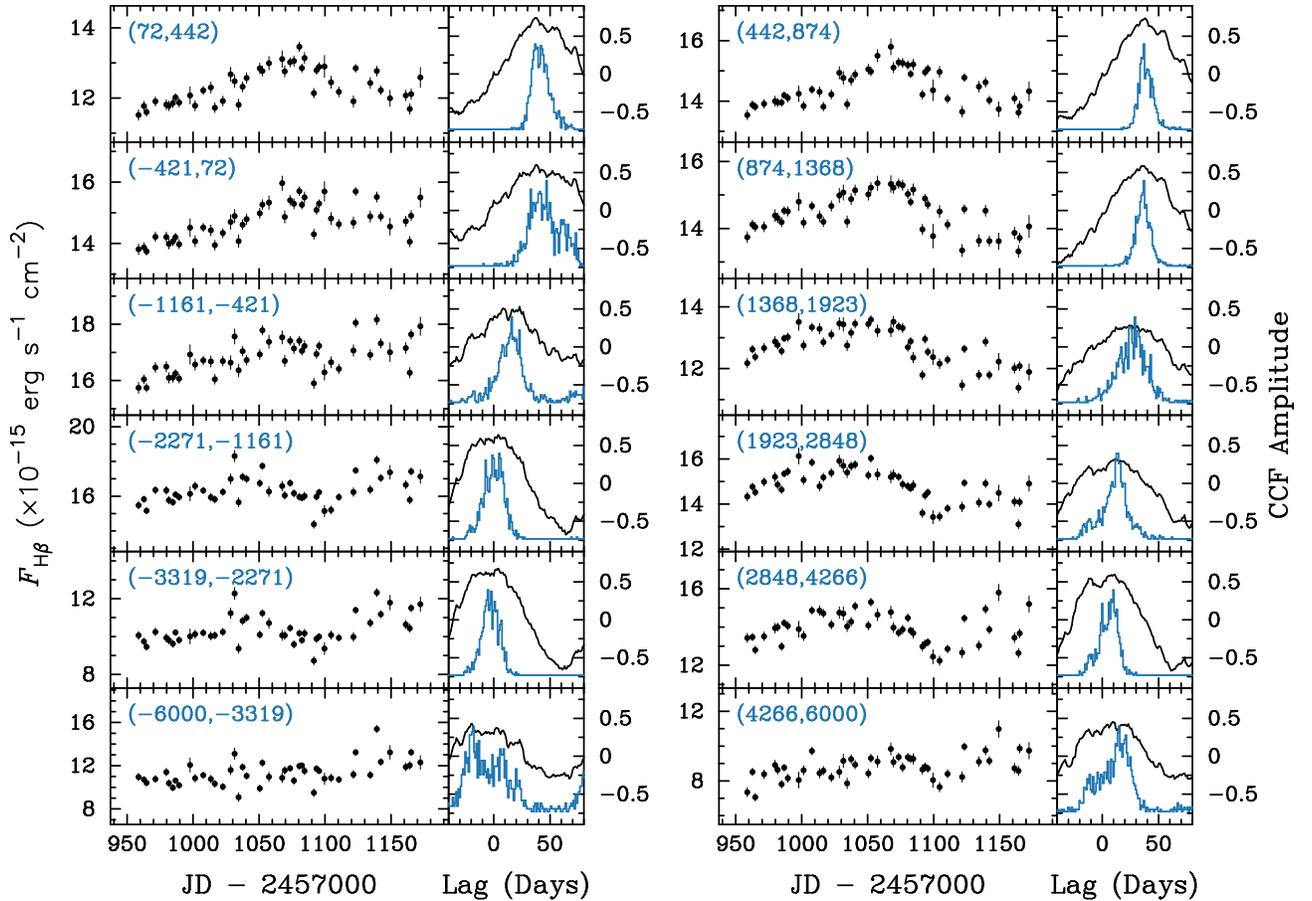}
  \caption{
  Light curves (dots with error bars), CCFs (black lines), and CCCDs (blue
  histograms) for all the velocity-space bins. The boundaries of each bin (in
  units of \kms) are written in the panel of each light curve. From
  bottom-left to top-left, and then from top-right to bottom-right, the
  velocity increases from negative (blueshift) to positive (redshift). 
  }
  \label{fig-vrlc17}
\end{figure*}

\begin{figure*}
  \centering
  \includegraphics[width=0.95\textwidth]{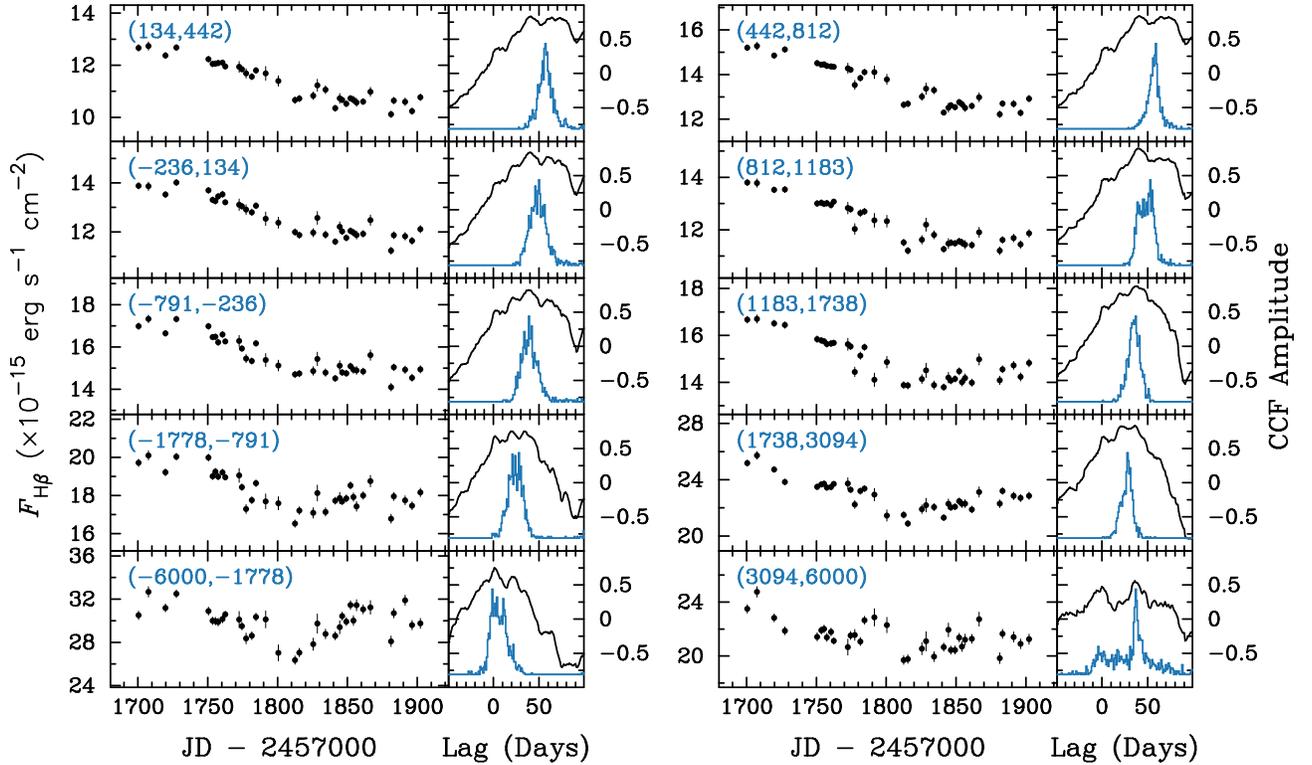}
  \caption{
  Light curves, CCFs, and CCCDs for all the velocity-space bins in 2019. The
  notations are as the same as those in Figure \ref{fig-vrlc17}.
  }
  \label{fig-vrlc19}
\end{figure*}

Figures \ref{fig-vrlc17} and \ref{fig-vrlc19} show the light curves
(black dots with error bars), and the corresponding CCFs (black curves) and
CCCDs (blue histograms) for all the velocity-space bins in 2017 and 2019,
respectively. The velocity range of each bin (in units of \kms) is written in
the panel of each light curve. It increases from negative (blueshift) to
positive (redshift) in a clockwise direction from bottom-left to bottom-right.
As in Section \ref{sec-lag}, a systematic error (not shown in the figure) has
been estimated and added before calculating the CCCD and the uncertainty of
the lag. The time lags (in the rest frame) and their uncertainties for all the
bins are plotted at corresponding flux-weighted velocities in the middle
panels of Figures \ref{fig-vrd17} and \ref{fig-vrd19}. The error bars in
the direction of velocity mark the widths of the bins. The blue and orange
horizontal solid lines show the lags of \hbv\ and \hbi\ listed in Table
\ref{tab-results}, respectively. The associated horizontal dashed lines are
one-sigma error above and below. We also plot the broad-\hb-only mean spectrum
(black histogram), the best-fit model (red curve) as the sum of \hbv\ (the
blue Gaussian) and \hbi\ (the orange Gaussian) in the top panels of
Figures \ref{fig-vrd17} and \ref{fig-vrd19}.

In 2017, the velocity-resolved delays are totally consistent with the
two-component scenario. In the bluest and reddest three bins at the wings
(between the two pairs of vertical blue dotted lines), the fluxes totally come
from \hbv, and the lags are roughly equal to that of \hbv. Note that the
bluest bin could be contaminated by \heii\ while the reddest three bins could
be influenced by \feii\ $\lambda$4924, from the uncertainties in fitting the
single-epoch spectra. On the other hand, for the four bins at the core
(between the orange vertical dotted lines), the variabilities are dominated by
that of \hbi. The lags are roughly constant at $\sim$35 days, which is
somewhat lower than the lag of \hbi\ because of the mixture of \hbv. For the
other two bins at the transition between the wings and the core, the lags also
transit from that at the wings to that at the core gradually as the fractions
of \hbi\ flux increase.

Three simple models with single kinematics are often used in the literature to
understand the results of velocity-resolved delays: a virialized disk, an
infall, and an outflow (see, e.g., Figure 10 in \citealt{bentz09}). However,
the velocity-resolved delays of PG 0026+129 in 2017 can not be interpreted by
any single one of these models. An virialized disk shows shorter lags at the
high-velocity wings, but not so discrete as we obtained here: those bins in
the two wings have lags of nearly zero, while the lags of bins for the line
core rise abruptly up to $\sim$35 days. The simplest interpretation is that
there are two distinct regions: a compact one emitting \hbv, plus another one
much far away for \hbi.

Another interesting result is the shape of the rms spectrum in 2017 shown in
the bottom panel of Figure \ref{fig-vrd17}. It shows a complex profile with
three peaks. When comparing with the mean spectrum (top panel), the core peak
at the velocity of $\sim$300 \kms\ matches \hbi, and the other two peaks
correspond to \hbv. Without the core component, the wings of the rms spectrum
show an asymmetric double-peaked profile with higher fluxes at the blue side.
Such a profile has been observed in many AGNs, and is associated with a
disk-like geometry \citep[e.g.,][]{storchi-bergmann17}. See Section
\ref{sec-hbv} below for more discussions.

In 2019, \hbi\ has an \fvar\ $\sim$3 times as large as that of \hbv\ (see
Table \ref{tab-results}). The rms spectrum (Figure \ref{fig-vrd19} bottom) is
dominated by the variability in \hbi, showing a strong core and weak wings,
which is much different in shape compared to that in 2017. Due to the low
fluxes at the wings in the rms spectrum, only the bluest and reddest bin
correspond to the \hbv-only region in the mean spectrum. The velocity-resolved
delays (Figure \ref{fig-vrd19} middle) still show a reliable lag consistent
with that of \hbv\ in the bluest bin, while the lag in the reddest bin is
highly uncertain (the CCF in this bin has two peaks, see the bottom-right
panel of Figure \ref{fig-vrlc19}). A possible reason is the contamination by
\feii\ $\lambda$4924, which would be severe in the event of the weak \hbv\
variability seen here. For bins other than the reddest and bluest, the lags
increase gradually towards the redshifted peak of the line, with increasing
flux fraction of \hbi. The velocity-resolved delays in 2019 are also
consistent with the two-component scenario, although the pattern is not as
discrete as that in 2017. The dominance of variability in \hbi\ over that in
\hbv\ weakens the contrast between the lags at the wings and the core. See
Section \ref{sec-hbi} below for more discussions on the much higher \fvar\ of
\hbi\ in 2019.

\section{Discussions}
\label{sec-discu}

\subsection{The Mass of the Central Black Hole}
\label{sec-mass}

The virial mass of the central black hole can be estimated from the
reverberation mapping measurements of the time lag $\tau$ and the
emission-line width $\Delta V$ as
\begin{equation}
  M_{\rm BH} = f \frac{c \tau \Delta V^2}{G}~,
\end{equation}
where $c$ is the speed of light, $G$ is the gravitational constant, and
$f$ is a virial factor counting for all other unknown effects including,
e.g., the geometry and kinematics of the emitting region.  In practice, $f$ is
obtained as an average for a sample of AGNs, by comparing the virial masses
with those given by other methods, e.g., the $M_{\rm BH}$--$\sigma_\ast$
relation \citep{onken04,grier13a}. The AGNs are classified into subsamples
according to, e.g., the properties of their bulges \citep{ho14}, to reduce the
uncertainty in the factor $f$. All the calibrations of $f$ in the
literature are done for the time lags and the line widths measured from the
total broad \hb\ line.

The line width can be measured as either FWHM or line dispersion (\sline),
in either the mean or rms spectrum. See \citet{peterson04} for a thorough
comparison of these methods. In principle, the rms spectrum
is preferred for providing the varying part of the emission line for which the
time lag is measured. But the rms spectrum usually has much lower S/N than the
mean spectrum, making the measurements more uncertain. In some cases, the rms
spectrum shows emission lines too weak to measure (e.g., PG 2130+099 in 2018;
Figure 2 of \citealt{hu20}), or dominated by other spectral components (e.g.,
the host galaxy, in MCG--6-30-15; \citealt{hu16}). The definition of FWHM is
somewhat arbitrary, especially for those complex multiple-peaked profiles (our
rms spectrum in 2017 as an example, bottom panel of Figure \ref{fig-vrd17}),
while \sline\ is well defined but sensitive to the subtraction of the
underlying continuum. Thus, in order to alleviate the uncertainty introduced
by the continuum subtraction and the contamination of \heii\ to the red wing
of \hb, we measure the FWHM and \sline\ in the broad-\hb-only mean and rms
spectra, which are generated after subtracting all other components given by
the spectral fitting, as for obtaining the velocity-resolved delays in Section
\ref{sec-vrd}. For FWHM, the method shown in Figure 1 of \citet{peterson04} is
adopted. The uncertainties are given by the standard deviations of the
values measured in Monte Carlo realizations (by bootstrap method) of the mean
and rms spectra.

\begin{deluxetable}{lr@{~$\pm$~}lr@{}lcr@{}l}
  \tablewidth{0pt}
  \tablecaption{Measurements for the Total Broad \hb\ Line
  \label{tab-hb}}
  \tablehead{
  \colhead{Method} & \multicolumn{2}{c}{Width} &
  \multicolumn{2}{c}{Virial Product} & \colhead{$f$} &
  \multicolumn{2}{c}{$M_{\rm BH}$}
  \\
  \colhead{} & \multicolumn{2}{c}{(\kms)} &
  \multicolumn{2}{c}{($\times10^7 M_\odot$)} & &
  \multicolumn{2}{c}{($\times10^7 M_\odot$)}
  \\
  \colhead{(1)} & \multicolumn{2}{c}{(2)} &
  \multicolumn{2}{c}{(3)} & \colhead{(4)} &
  \multicolumn{2}{c}{(5)}
  }
  \startdata
  \multicolumn{8}{c}{2017}
  \\ \tableline
  mean, FWHM &
  3374 & 27  & 2.59 & $_{-1.73}^{+1.65}$ & 1.3 & 3.37 & $_{-2.25}^{+2.15}$
  \\
  mean, \sline &
  2274 & 4   & 1.18 & $_{-0.78}^{+0.75}$ & 5.6 & 6.60 & $_{-4.39}^{+4.20}$
  \\
  rms, FWHM &
  2735 & 578 & 1.70 & $_{-1.34}^{+1.30}$ & 1.5 & 2.56 & $_{-2.02}^{+1.95}$
  \\
  rms, \sline &
  2446 & 88  & 1.36 & $_{-0.91}^{+0.87}$ & 6.3 & 8.59 & $_{-5.75}^{+5.50}$
  \\ \tableline
  \multicolumn{8}{c}{2019}
  \\ \tableline
  mean, FWHM &
  3198 & 21  & 5.53 & $_{-1.21}^{+1.00}$ & 1.3 & 7.19 & $_{-1.57}^{+1.30}$
  \\
  mean, \sline &
  2315 & 4   & 2.90 & $_{-0.63}^{+0.52}$ & 5.6 & 16.2 & $_{-3.5}^{+2.9}$
  \\
  rms, FWHM &
  1902 & 114 & 1.95 & $_{-0.49}^{+0.42}$ & 1.5 & 2.93 & $_{-0.73}^{+0.63}$
  \\
  rms, \sline &
  1901 & 97  & 1.95 & $_{-0.47}^{+0.40}$ & 6.3 & 12.3 & $_{-3.0}^{+2.5}$
  \\
  \enddata
  \tablecomments{
  Widths of the total broad \hb\ line (column 2) measured by different methods
  (column 1) in years 2017 and 2019. The instrumental broadening has been
  corrected. Column (3) lists the virial products. Column (5) lists the masses
  of the central black hole estimated using the virial factors $f$ (column 4)
  correspond to different width measurements given by \citet{ho14}. The
  uncertainty in $f$ has not been included.
  }
\end{deluxetable}

Table \ref{tab-hb} gives the widths (column 2) of the total broad \hb\
measured by different methods (column 1) in years 2017 and 2019. It can be
seen that the shapes of the mean spectra in the two years are almost the same
(compare the top panels of Figures \ref{fig-vrd17} and \ref{fig-vrd19}). The
changes in the widths presented by both FWHM and \sline\ are less than 5\%.
With a time lag in 2019 more than twice as long as that in 2017, the virial
products (VPs, defined as $c \tau {\rm FWHM}^2 / G$ or $c \tau \sigma_{\rm
line}^2 / G$ for FHWM or \sline, respectively; column 3) in 2019 are also
more than twice as large. On the other hand, the shapes of the rms spectra
change significantly between the two years, and thus the widths as well. Both
FWHM and \sline\ are much smaller in 2019, yielding more consistent VPs between
the two years than by mean spectra. Especially, when FWHM in the rms spectrum
is used, the difference in VPs between the two years is $\lesssim$15\%. As
mentioned in Section \ref{sec-lag}, the lag of \hbt\ is roughly the
varying-flux-weighted average of the \hbv\ and \hbi\ lags. The large
increasing of \hbi\ \fvar\ in 2019 accordingly strengthens \hbi\ in the rms
spectrum, and thus decreases the line width. The dramatic changes in the time
lags and the rms spectra between the two years are both caused by the
different behaviors of the two \hb\ components. And the FWHM in the rms
spectrum is preferred for line width measurements, as in this case it yields
the most consistent VPs between the two years. 

Column (4) of Table \ref{tab-hb} lists the virial factors $f$ corresponding to
different line width measurements from \citet{ho14} for a classical bulge (see
\citealt{ho14} for a discussion on the bulge type of PG 0026+129), and column
(5) lists the resultant black hole masses. Note that the masses given by
\sline\ are several times higher than those given by FWHMs for the extremely
small values of FWHM/\sline\ of this target (see Figure 9 of
\citealt{peterson14} for a comparison). Such a small FWHM/\sline\ ($\sim$1)
in the rms spectra indicates that PG 0026+129 has wings much more variable
than for most other objects. And the value of $f$ in the table given as the
mean in a sample is very probably unsuitable in this extreme case. The direct
modeling method \citep{pancoast11} could provide an estimate of the black hole
mass without the assumption of $f$, but is out of the scope of this work.
Therefore, considering that the masses given by the FWHMs in the rms spectra
have the best consistency between the two years, we obtain the mass of the
central black hole in PG 0026+129 as the weighted mean of the values given by
this method: $M_{\rm BH} = 2.89_{-0.69}^{+0.60} \times10^7 M_{\odot}$.

Previous estimations of the black hole mass of PG 0026+129 were based on the
time lag measured by \citet{kaspi00}, and were several times larger than
the results here if the same method for line width measurement and $f$ are
used. E.g., the VP given by \sline\ in the rms spectra remeasured by
\citet{peterson04} is $7.14\pm1.74 \times10^7 M_{\odot}$, $\sim$4--5 times as
large as our results by the same method. Possibly the time lag was
overestimated in \citet{kaspi00} for undersampling \citep{grier08}, but there
is no reliable black hole measurement by other methods for a comparison. The
stellar velocity dispersion for PG 0026+129 has not been successfully measured
in previous studies \citep{grier13a}, and the masses of its host galaxy or
bulge are also largely uncertain. \citet{ho14} gave a rather large bulge mass
of $2.1 \times 10^{11} M_{\odot}$, based on the $R$-band magnitude. However,
\citet{bentz18} derived a much smaller mass of $1.7 \times 10^{10} M_{\odot}$,
by estimating the mass-to-light ratio using the $V-H$ color. Using their
Equation (3) for the $M_{\rm BH}$--$M_{bulge}$ relation, the expected black
hole mass is only $1.9 \times 10^7 M_{\odot}$. Better observations of the host
galaxy, both multi-band photometry and spectroscopy, are needed for a reliable
bulge mass estimation.

Comparing with the total \hb\ line, \hbi\ or \hbv\ can be potentially
better for the virial mass estimation, because each of these components is
hopefully less complex in geometry than the total line. The VPs given by the
lag of \hbi\ and its FWHM in the mean spectrum (see Table \ref{tab-results})
are $3.27 \times 10^7$ and $4.52 \times 10^7 M_{\odot}$ in years 2017 and
2019, respectively. These values are consistent with those given by the total
line with the same method (FWHM in the mean spectrum), and the difference
between the two years is smaller. The widths of the two components in the
rms spectrum are presumably more suited for the mass estimation than the
widths in the mean spectrum, as the former represent the varying part of each
component. However, the decomposition of the two components in the rms
spectrum is not so straightforward, due to its complex shape. On the other
hand, the factor $f$ for each single component is totally unknown so far.

\subsection{No Long-Term Variation in the Broad \hb\ Profile}
\label{sec-asymm}

As mentioned in Section \ref{sec-fit}, both \hbv\ and \hbi\ are redshifted
with respect to the narrow lines, by velocities of $\gtrsim$400 \kms\ measured
from the mean spectrum. Because of the relatively low spectral resolution of
our spectra, it is not reliable to study the variations in the velocity shifts
of the two components between different epochs during our campaign. But it is
interesting to compare the \hb\ profile in our spectra with those in
\citet{boroson92} and \citet{kaspi00} for long-term variations in years.

\begin{figure}
  \centering
  \includegraphics[angle=-90,width=0.475\textwidth]{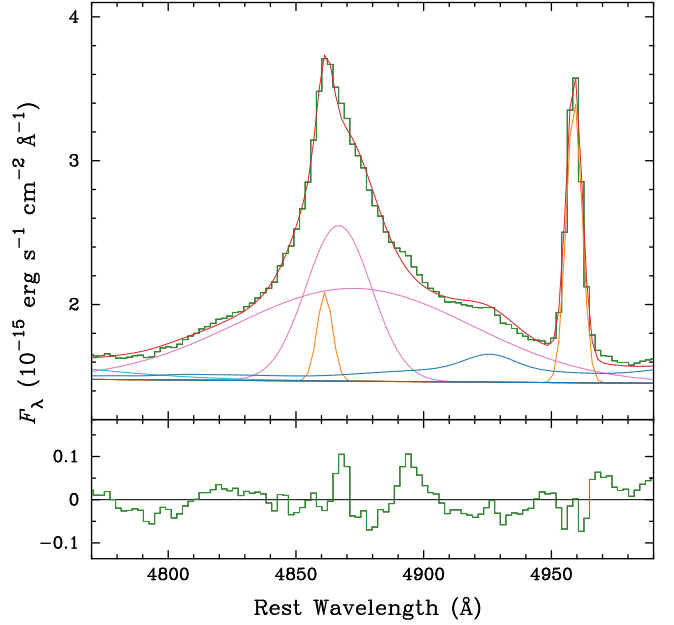}
  \caption{
  Fit of the archived spectrum in \citet{boroson92}. The components involved
  in the fitting and the notations are as the same as those in Figure
  \ref{fig-spec}. Note that their spectrum has higher spectral resolution and
  shows the asymmetry in the \hb\ profile more clearly than our spectra.
  }
  \label{fig-bg92spec}
\end{figure}

PG 0026+129 was observed in Oct 1990 with better spectral resolution
($\sim$360 \kms) by \citet{boroson92}, and the spectrum is archived in the
NED. As shown in Figure \ref{fig-bg92spec}, we fit the \hb\ line with three
Gaussians: a narrow component (in orange) which is forced to have the same
velocity shift and width as the \oiii\ lines, and the other two Gaussians (in
magenta) represent \hbv\ and \hbi. The best-fit FWHMs (after instrumental
broadening correction) and velocity shifts are $\sim$6710 \kms\ and $\sim$700
\kms\ for \hbv, $\sim$1820 \kms\ and $\sim$340 \kms\ for \hbi, respectively.
Note that the spectral shape of the archived spectrum is not well calibrated
(the fluxes redward of the rest-frame 5100 \AA\ are lower than those of a
power law extrapolated from the blueward part of the spectrum), so the
measurements of broad \hb\ components, especially \hbv, are influenced by the
uncertain continuum level. However, with their high spectral resolution,
the \hb\ profile of \citet{boroson92} clearly shows: 1) a narrow peak at zero
velocity shift, indicating that the \oiii\ lines are not blueshifted with
respect to the low-ionizing narrow lines, and are thus appropriate for
defining the systematic redshift of the object; 2) significant asymmetry,
which is caused by the redshifted broad components, especially \hbi.

\begin{figure}
  \centering
  \includegraphics[width=0.475\textwidth]{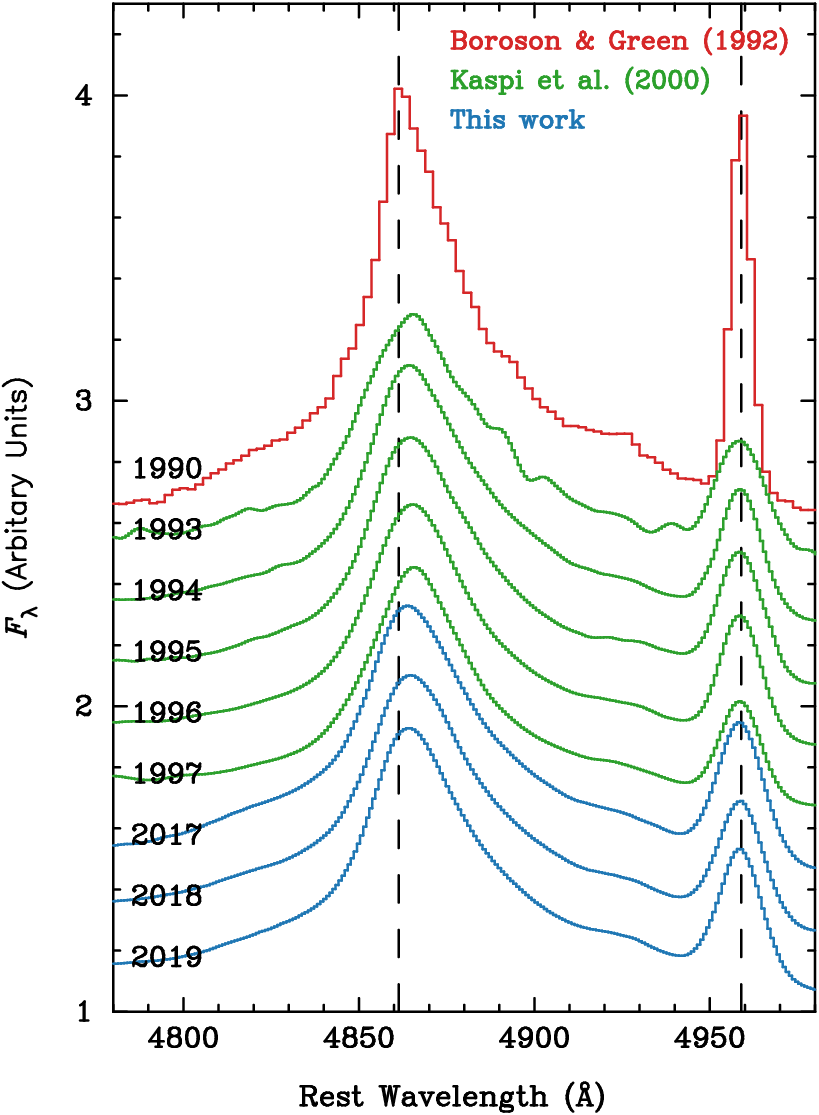}
  \caption{
  \hb\ profiles in different years from \citet{boroson92} (in red),
  \citet{kaspi00} (in green), and this work (in blue). The spectra are
  normalized and vertically shifted for clarity. The years when the spectra
  were observed are noted. The two vertical dashed lines mark the positions of
  zero shifts for \hb\ $\lambda$4861 and \oiii\ $\lambda$4959.
  }
  \label{fig-hbcomp}
\end{figure}

The spectra in the \citet{kaspi00} campaign%
\footnote{\url{http://wise-obs.tau.ac.il/~shai/PG/}}
also have low spectral resolution just comparable to that in this work (wide
slits are usually used in reverberation mapping observations for good flux
calibration). We generate mean spectra for the years 1993 to 1997, in which
more than five epochs were observed. Figure \ref{fig-hbcomp} plots the mean
spectra in these five years (green), along with the spectrum of
\citet{boroson92} (red) and the mean spectra of the three years in our
campaign (blue). The spectra are normalized by their flux at rest-frame
5100\AA, and shifted vertically for clarity. The two vertical dashed lines
mark the positions of \hb\ $\lambda$4861 and \oiii\ $\lambda$4959 with zero
velocity shifts. See the widths of \oiii\ $\lambda$4959 for a comparison of
the spectral resolutions. For the spectra other than that of
\citet{boroson92}, the peaks of narrow \hb\ lines are smoothed by the large
instrumental broadening and can no longer be distinguished. But the redshifts
of the broad line core, consisting mainly \hbi, are clear in all the spectra.
We decomposed the profiles and attempted to compare the velocity shifts of
\hbi\ between different years. No reliable variation was confirmed, presumably
because of the uncertainty due to the unresolved narrow component.

In summary, in view of the limited quality of the data, there is no evidence
of significant change in the \hb\ profile in these years. Most probably, the
redshift of \hbi\ remains roughly the same in nearly 30 years.

\subsection{Intermediate-Width Component: an Infall?}
\label{sec-hbi}

The \feii\ emission lines of PG 0026+129 have similar widths to \hbi, and are
also redshifted, suggesting that both \hbi\ and \feii\ are emitted from an
intermediate-line region. \citet{hu08b} measured the velocity shifts
of \feii\ emission in a large sample of quasars, and found that \feii\ are
systematically redshifted (see \citealt{sulentic12,hu12,bon18,le19} for more
discussions). The inverse correlation between the shift and the Eddington
ratio in \citet{hu08b} indicates that these intermediate-width lines
originate from an infall, because the radiation pressure increases with
higher Eddington ratio and decelerates the infall more \citep{ferland09}. A
similar inverse correlation between the velocity shift and the continuum
flux should also exists in multi-epoch observations of a single object, as
expected for an infall. However, as shown in Section \ref{sec-asymm}, the
current data allow no such an exploration. 

The size of this intermediate-line region could be constrained by the time
lags of \hbi\ we measured in the two years, 43.4 and 60.0 light days
respectively. Adopting the mean fluxes of AGN continuum at 5100\AA\
given by our spectral fitting ($2.74 \times 10^{-15}$ \ergscma\ in 2017
and $3.03 \times 10^{-15}$ \ergscma\ in 2019) and a luminosity distance%
\footnote{Based on $z$ = 0.1454 and cosmological parameters of $H_0=72~{\rm
km~s^{-1}}$ Mpc$^{-1}$, $\Omega_{m}=0.3$, and $\Omega_\Lambda=0.7$.
}
of 670 Mpc, we obtain spectral luminosities of $\lambda L_{\lambda}$(5100 \AA)
= $7.5 \times 10^{44}$ erg s$^{-1}$ and $8.3 \times 10^{44}$ erg s$^{-1}$,
respectively. Then the BLR radius--luminosity relation of \citet{bentz13}
predicts radii of 98 and 104 light days, but for the whole \hb\ line. Our
measurement of \hbi\ is roughly a half of the value predicted, and that of
\hbt\ is only 1/8 to 1/4, by including the rather compact \hbv. Note that
shorter time lags than implied by that relation have been observed in many
objects \citep{du15,grier17b}. Especially for SEAMBH, time lags shortened by a
factor of two are common and interpreted as a consequence of the
self-shadowing effects \citep{wang14,du18b}. Adopting the black hole mass
estimated in Section \ref{sec-mass}, the dimensionless accretion rates defined
by equation (2) in \citet{du15} have rather high values of 76 and 88, for the
two years, indicating a SEAMBH in PG 0026+129. Thus, the small size of the
intermediate-line region we measured is consistent with those in other
SEAMBHs. However, such an extremely compact very-broad-line region that emits
nearly 3/4 of the total fluxes has not been seen before, and most probably has
a different origin.

The lag of \hbi\ is $\sim$50\% longer in 2019 than it was in 2017, while the
AGN continuum is only 10\% more luminous. In addition, \hbi\ is much more
variable (nearly 70\% larger \fvar) in 2019, while the variability amplitude
of the continuum is mildly smaller. Another parameter worth noting here
is the continuum variability timescale, which is apparently longer in 2019
(see the top panel of Figure \ref{fig-lcfit}). It is possible that the \hbi\
region is more extended than what we measured here using time lags. As shown
by the photoionization calculations and light-curve simulations in
\citet{goad14}, the measured time lag and also the line responsivity will be
reduced if the continuum varies faster than the maximum lag corresponding to
the outer boundary of the emission-line region, because of the geometric
dilution (see their Figure 9). Our results of the lags, variability amplitudes
and timescales in the two years match this dilution effect qualitatively.

\subsection{Very Broad Component: the Accretion Disk?}
\label{sec-hbv}

As shown in Section \ref{sec-vrd}, the wings of the rms spectrum in 2017,
corresponding to the velocity range of \hbv, show an asymmetric double-peaked
profile. But in the mean and single-epoch spectra, \hbv\ is fitted well by a
single Gaussian (see Figure \ref{fig-spec} and the top panel of Figure
\ref{fig-vrd17}). Different \hb\ profiles in the mean and rms spectra are
commonly seen in previous reverberation mapping campaigns, and interpreted as
only a part of the total line fluxes are variable to respond to the continuum
variations \citep[e.g.,][]{peterson04}.

Double-peaked broad emission lines have been seen in many objects, including
both low-luminosity AGNs \citep[e.g.,][]{ho00,shields00,bianch19} and Seyfert
1 galaxies \citep[e.g.,][]{storchi-bergmann17}. The line is believed to be
generated in the accretion disk itself \citep{strateva03}, or the inner part
of the BLR which is just the outward extension of the accretion disk
\citep{storchi-bergmann17}. The near-zero time lag between \hbv\ and the
optical continuum we measured in PG 0026+129 suggests that this
very-broad-line region has to be tightly associated with the accretion disk.
This very inner part of the BLR may just originate from the surface of the
accretion disk, as suggested by some authors
\citep[e.g.,][]{dumont90,czerny11}.

Following \citet{bianch19}, we fit the asymmetric double-peaked wings of the
rms spectrum with the \textsc{kerrdisk} model developed by
\citet{brenneman06}, which simulates the broad line emitted from an accretion
disk system. The fitting is performed using \textsc{xspec} 12.10.1
\citep{arnaud96}, and the results are shown in the bottom panel of Figure
\ref{fig-vrd17}. The \textsc{kerrdisk} model is convolved with a Gaussian
smoothing (in blue, after convolving), and an additional Gaussian line (in
orange) is added as the intermediate-width component. The best fit (in red)
constrains following parameters for the disk as: the emissivity index is
$2.16_{-0.12}^{+0.17}$; the inclination angle to the line of sight is
$22.0\degr_{-0.2}^{+0.5}$; the inner and outer radii are $152_{-15}^{+15}$ and
$1389_{-137}^{+924}$, respectively, in units of gravitational radius defined
as $GM/c^2$. The dimensionless spin of the black hole is not well constrained
as $0.46_{-0.17}^{+0.06}$, for the much larger inner radius than the
marginally stable radius. Adopting $M_{\rm BH} = 2.89 \times10^7 M_{\odot}$
estimated in Section \ref{sec-mass}, the inner and outer radii are 0.25 and
2.29, in units of light days. For comparison, on a standard centrally
illuminated thin accretion disk, the characteristic radius emitting at
5100\AA\ given by equation (1) of \citet{edelson19} is 1.3 light days in the
flux-weighted case, adopting an Eddington ratio of 1.9 in 2017 (by $L_{\rm
bol} = 9 \lambda L_\lambda$(5100 \AA)). So the size of the disk given by
modeling the double-peaked profile is consistent with the zero time lag of
\hbv\ we measured, supporting the accretion disk origin of the very broad
component.

On the other hand, the diffuse continuum emission from the line-emitting
clouds is unavoidable and contributes rather significantly to the observed
optical continuum in the calculations of several models of the BLR
\citep[e.g.,][]{korista01,chelouche19,netzer20}. Thus the time lag of the
optical continuum with respect to the UV continuum could be longer than that
given by the illuminated disk model by a factor of a few times for the
contribution of this non-disk continuum
\citep[e.g.,][]{lawther18,chelouche19,korista19}. In this case, the variations
in the \hbv\ emission from the disk fitted above will be a few days leading
those of the 5100 \AA\ continuum. However, the quality of our data set, mainly
the sampling interval, allows no reliable determination of negative lags of a
few days. Interestingly, in the model of \citet{chelouche19}, the non-disk
continuum emission is emitted by the gas launched from the accretion disk,
which may correspond to the region responsible for \hbv\ in terms of size,
although in their model high gas density suppressed Balmer lines by
collisional de-excitation \citep{baskin14}.

The photoionized accretion disk model in \citet{dumont90} produces copious
Balmer lines but collisionally suppressed \lya\ \citep{rokaki92}, as observed
in Arp 102B. The double-peaked components are strong in Balmer lines, and can
be fitted by a disk with similar size in units of gravitational radius
\citep{halpern96} as that given for PG 0026+129 above. But \lya\ shows no
such a disk component in Arp 102B (Figure 3 of \citealt{halpern96}). By
contrast, the profile of \lya\ line in PG 0026+129 (see \textit{Hubble Space
Telescope}/Faint Object Spectrograph spectrum collected by
\citealt{bechtold02}) shows strong wings, even broader than \hbv, indicating
that the collisional de-excitations are not dominant in the \hbv-emitting
clouds in PG 0026+129. Detailed photoionization modeling, which is beyond the
scope of this paper, may reveal why PG 0026+129 is unique (so far) in having
such strong line emission emitted so close to the ionizing source.

In 2019, while the mean spectrum retains the same strong \hbv, the rms
spectrum shows no double-peaked wings as clearly as in 2017, as a result of
the more variable \hbi. But relatively weak, very-broad wings are evident.
Thus, the rare existences of outstanding \hbv\ in both the velocity-resolved
delays and the rms spectrum in the literature do not necessarily mean that
such an \hbv\ is unique for PG 0026+129. It is possible that compact disk-like
\hbv\ also exist in other AGNs, but not as strong and variable as that of PG
0026+129 in 2017, or just hide beneath the other more variable parts of the
BLR as in the case of 2019 here. High quality velocity-resolved delay
measurements would hopefully reveal this kind of \hbv\ in more AGNs, with fast
driving continuum variations.

\section{Summary}
\label{sec-sum}

We performed a new reverberation mapping campaign of the quasar PG 0026+129
using the CAHA 2.2m telescope lasting three years from 2017 to 2019. In the
first and third years, the object has sizable variations in the
continuum fluxes, and significant reverberations of broad \heii\ and \hb\
emission lines are detected. The spectral decomposition and time series
analysis show strong evidence that two kinematically and geometrically
distinct \hb-emitting regions exist. The main results can be summarized as
follows.

\begin{enumerate}
  \item The broad \hb\ emission line can be decomposed to two components: a
    very broad \hbv\ with a FWHM of 7570$\pm$83 \kms, and another
    intermediate-width \hbi\ with a FWHM of 1964$\pm$18 \kms. Both
    components show significant reverberations to the continuum variations.
    The time lags (in the rest frame) are $-1.9_{-5.5}^{+9.3}$ and
    $-1.1_{-2.2}^{+12.9}$ days for \hbv, $43.4_{-1.5}^{+4.1}$ and
    $60.0_{-11.0}^{+5.9}$ days for \hbi, in 2017 and 2019, respectively.
  \item The velocity-resolved delays are roughly zero at the \hb\ wings and
    $\sim$30--50 days at the core, with no gradual transition between these
    regimes, supporting the existence of two distinct broad \hb\ components.
  \item \hbi\ and \feii\ emission have similar line widths, and both are
    redshifted, indicating that they both originate from an intermediate-width
    line region which could be an infall.
  \item In 2017, a reliable lag of $-1.4_{-6.9}^{+4.9}$ days for the
    broad \heii\ line is also detected. \hbv\ and \heii\ have similar line
    widths and time lags. We suggest that both of them are emitted from a
    region associated with the accretion disk, because: 1) the lags of close
    to zero indicate that the region has a size comparable to that of the part
    of the accretion disk emitting the optical continuum; 2) the rms spectrum
    of \hbv\ shows an asymmetric double-peaked profile which suggests a
    disk-like structure.
  \item Combining the time lags for the total \hb\ broad line and the
    FWHMs in the rms spectra yields the mass of the central black hole with
    the best consistency between the two years. The weighted mean $M_{\rm BH}
    = 2.89_{-0.69}^{+0.60} \times10^7 M_{\odot}$ is adopted, assuming a virial
    factor of 1.5.
\end{enumerate}

\acknowledgments
We acknowledge the support of the staff of the CAHA 2.2m telescope. This work
is based on observations collected at the Centro Astron\'omico Hispanoen
Andaluc\'ia (CAHA) at Calar Alto, operated jointly by the Andalusian
Universities and the Instituto de Astrof\'isica de Andaluc\'ia (CSIC). This
research is supported by the National Key R\&D Program of China
(2016YFA0400701, 2016YFA0400702), by the National Science Foundation of China
(11721303, 11773029, 11833008, 11873048, 11922304, 11973029, 11991051,
11991052, 11991054), by the Key Research Program of Frontier Sciences of the
Chinese Academy of Sciences (CAS; QYZDJ-SSW-SLH007), by the CAS Key Research
Program (KJZDEW-M06), and by the Strategic Priority Research Program of the
CAS (XDB23000000, XDB23010400). JA acknowledges financial support from the
State Agency for Research of the Spanish MCIU through the ``Center of
Excellence Severo Ochoa'' award to the Instituto de Astrof\'isica de
Andaluc\'ia (SEV-2017-0709).

\appendix

\section{Results from the Entire Data Set of Three Years}
\label{sec-3y}

This Appendix presents the results of the time-series analysis performed on
the entire data set of the three years. In general, the time lags measured
from data sets with multiple observing seasons should be treated with caution,
because of the gaps between the observing seasons and possible different
long-term trends in the light curves. For comparison, the \textsc{javelin}
software \citep{zu11} was also used for the three-year data set. By assuming a
damped-random-walk model and a top-hat transfer function, \textsc{javelin}
treats the seasonal gaps in a more sophisticated manner than ICCF. The results
from \textsc{javelin} and ICCF are generally consistent, and the numbers list
below are given by ICCF.

\begin{figure}
  \centering
  \includegraphics[width=0.475\textwidth]{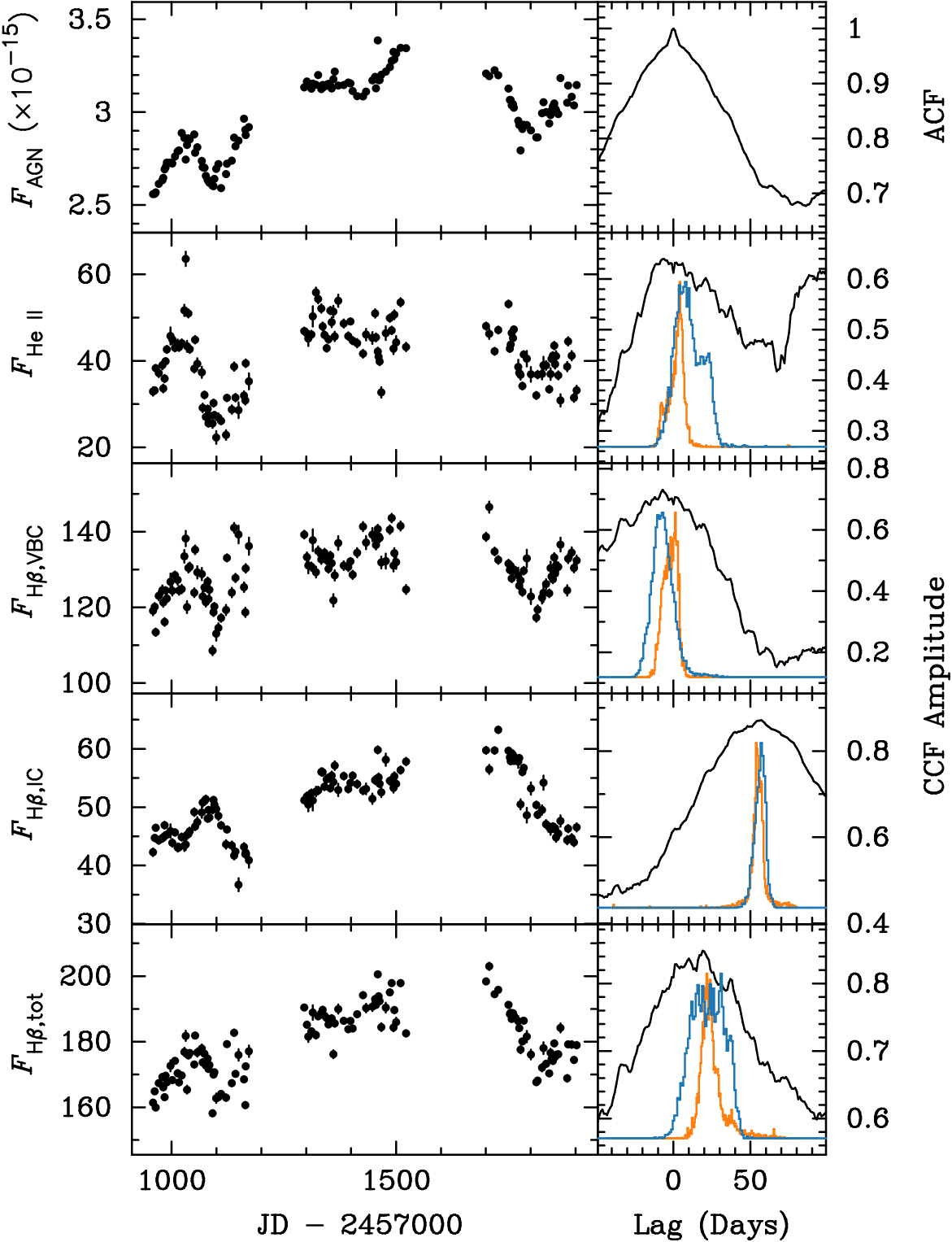}
  \caption{
  Light curves and CCF analysis results for the entire three years. The units
  and notations are as the same as those in Figure \ref{fig-ccf17}, while the
  additional orange histograms are the \textsc{javelin} posterior
  distributions of the lags. 
  }
  \label{fig-ccf3y}
\end{figure}

Figure \ref{fig-ccf3y} shows the light curves, CCFs, CCCDs, and the
\textsc{javelin} posterior distributions of lags for the broad emission lines
and components. The time lags are $-6.1_{-5.6}^{+5.9}$ and
$49.6_{-3.8}^{+2.3}$ days in the rest frame for \hbv\ and \hbi, respectively.
For the total \hb, the measured time lag is $13.4_{-3.7}^{+15.9}$ days. The
light curves of the two \hb\ components and the total \hb\ show no obvious
long-term trend differing from the AGN continuum.

\begin{figure}
  \centering
  \includegraphics[width=0.45\textwidth]{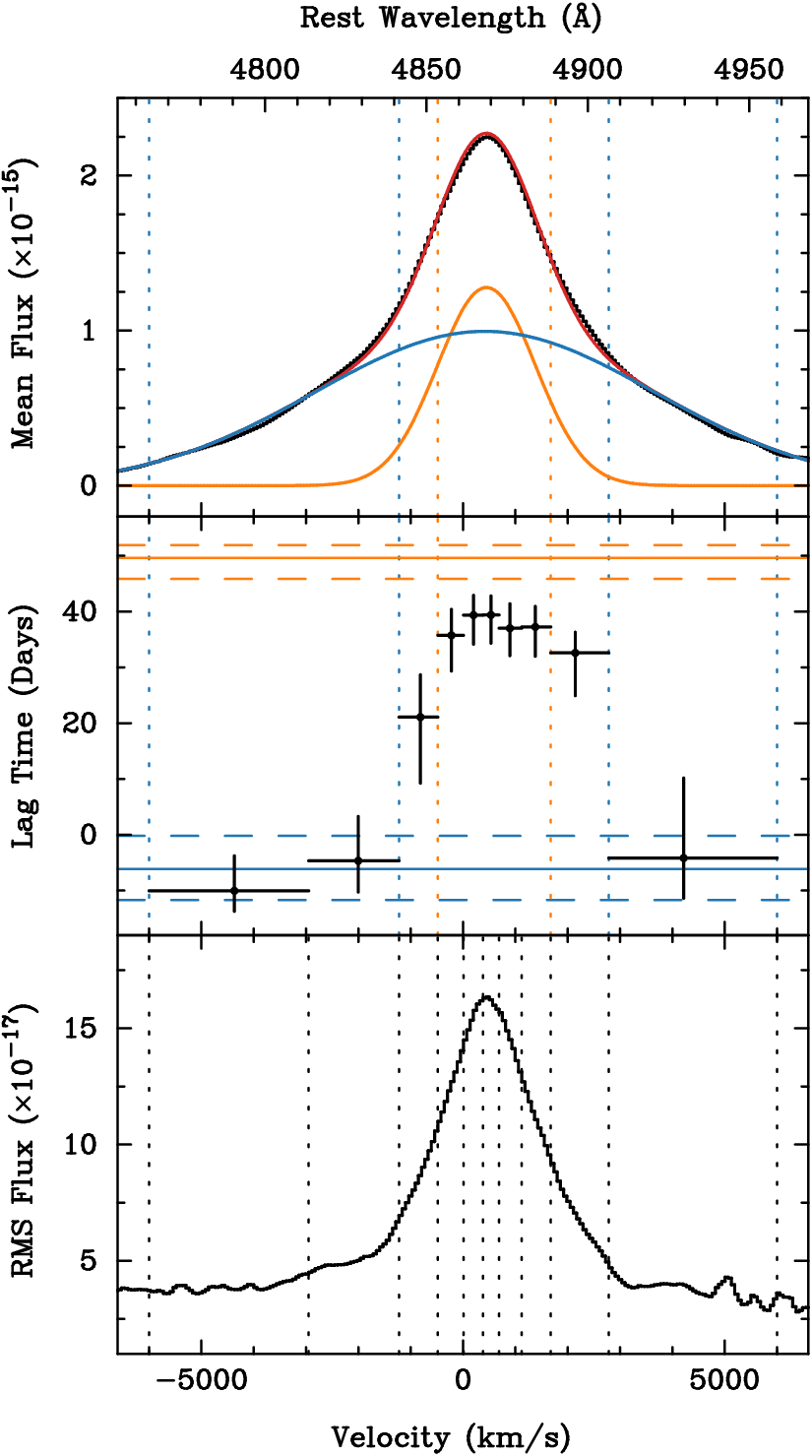}
  \caption{
  The broad-\hb-only mean spectrum (top), the velocity-resolved delays
  (middle), and  the broad-\hb-only rms spectrum (bottom) for the entire three
  years. The notations are as the same as those in Figure \ref{fig-vrd17}.
  Note that the measurements in the narrow velocity bins around the core are
  not independent due to the instrument broadening.
  }
  \label{fig-vrd3y}
\end{figure}

\begin{figure*}
  \centering
  \includegraphics[width=0.95\textwidth]{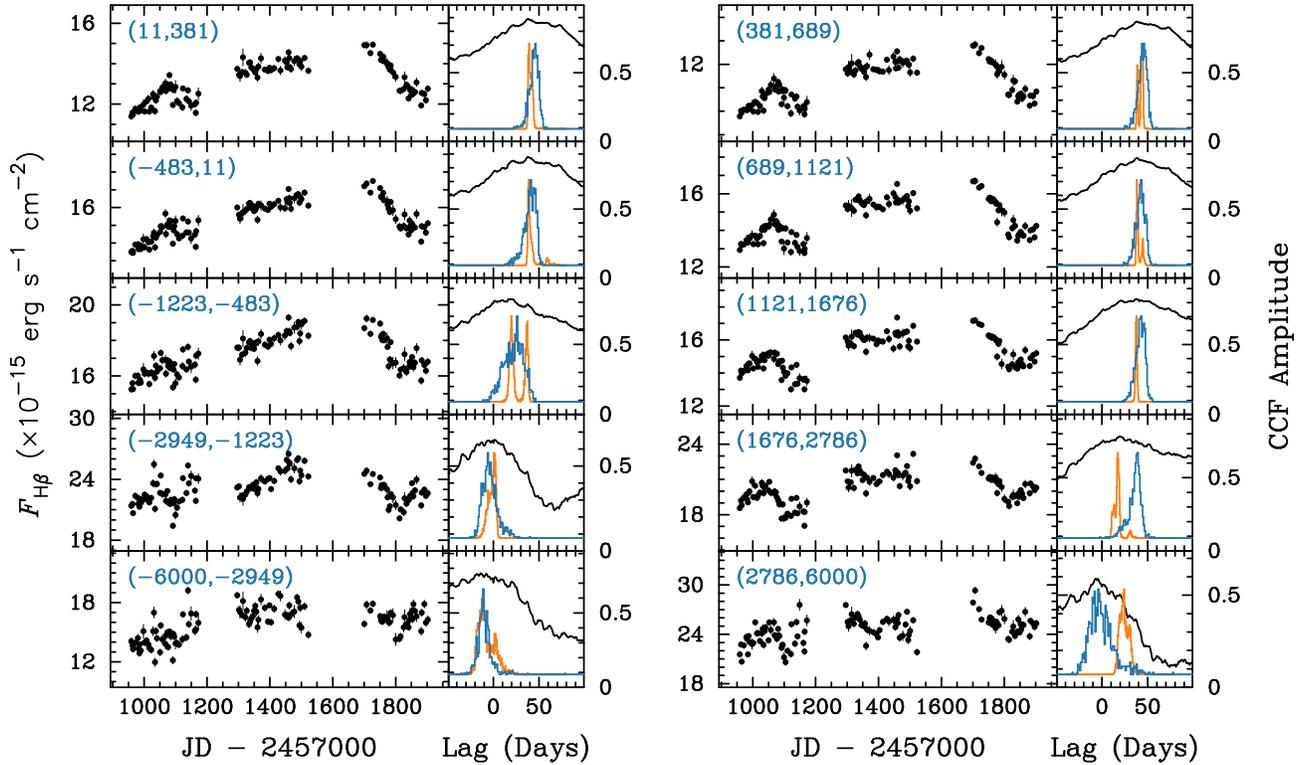}
  \caption{
  Light curves, CCFs, and CCCDs for all the velocity-space bins for the entire
  three years. The notations are as the same as those in Figure
  \ref{fig-vrlc17}, while the additional orange histograms are the
  \textsc{javelin} posterior distributions of the lags.
  }
  \label{fig-vrlc3y}
\end{figure*}

Figure \ref{fig-vrd3y} shows the broad-\hb-only mean spectrum and the
decomposition, the velocity-resolved delays, and the rms spectrum. Figure
\ref{fig-vrlc3y} shows the light curves, CCF analysis, and \textsc{javelin}
posterior distributions for each velocity bin. The differences between the
lags from ICCF and \textsc{javelin} in the bluest two bins indicate the
measurements in these bins have large uncertainties as in the single-year data
set mentioned in Section \ref{sec-vrd}, which may be influenced by the
inaccurate decomposition of \feii\ $\lambda$4924 in the spectral fitting. The
velocity-resolved delays show an obvious discrete structure consistent with
the two-component scenario: the lags in wing bins equal to that of \hbv, while
the lags in the core bins jump up close to that of \hbi.

\section{Spectra obtained with the Sutherland 1.9 m telescope}
\label{sec-saao}

During the campaign, PG 0026+129 was also observed by the Sutherland 1.9 m
telescope at the South African Astronomical Observatory. Spectra were taken
with the 600 lines mm$^{-1}$ grating and $4\farcs0$ slit for several epochs
(see \citealt{winkler17} and also \citealt{hu20} for more details on the
observations and data reduction). The spectral resolution is $\sim$340 \kms\
estimated by the FWHM of the sky line, better than that of our CAHA spectra.
In addition, \oii\ $\lambda$3727 is covered. Thus, we present a Sutherland
spectrum here to investigate whether \oiii\ is blue-shifted with respect to
the low-ionization lines.

\begin{figure}
  \centering
  \includegraphics[angle=-90,width=0.475\textwidth]{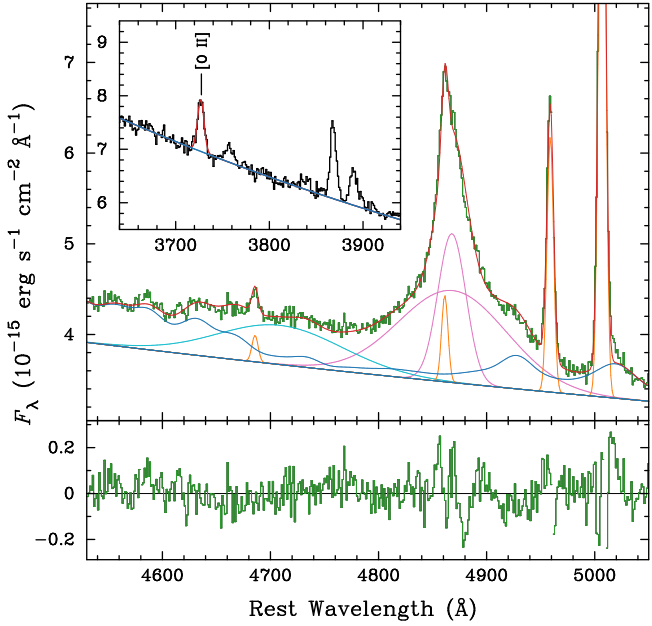}
  \caption{
  Fit of a spectrum obtained with the Sutherland 1.9 m telescope on JD
  2458322. The spectrum has been deredshifted according to the \oiii\ lines.
  The components involved in the fitting and the notations are as the same as
  those in Figures \ref{fig-spec} and \ref{fig-bg92spec}. The inserted plot
  shows the fit by a Gaussian (red) above a local continuum (blue) to \oii\
  $\lambda$3727, whose position of zero velocity shift is marked. Note that
  the \oiii\ lines are not blue-shifted with respect to either the \hb\ peak
  or \oii.
  }
  \label{fig-saao}
\end{figure}

Figure \ref{fig-saao} shows a spectrum taken on JD 2458322 in July 2018. It is
well fitted by the same spectral components described in Section
\ref{sec-fit}. The FWHMs (after instrumental broadening correction) and the
velocity shifts (with respect to \oiii) are 7225 \kms\ and 449 \kms\ for \hbv,
1847 \kms\ and 403 \kms\ for \hbi, respectively. \oii\ $\lambda$3727 is
clearly detected as shown in the inserted plot. A simple fit to the doublet
with a single Gaussian above a locally defined continuum yields a blueshift of
48 \kms\ with respect to the \oiii\ lines. The uncertainty in the measurement
of the \oii\ shift introduced by the unknown line ratio of the doublet has to
be smaller than the pair separation (2.78 \AA\ or 224 \kms). Thus, the
$\gtrsim$400 \kms\ redshifts of the two broad \hb\ components can't be
interpreted as \oiii\ being blue-shifted in this object. In addition, the peak
of the \hb\ profile is well fitted by the narrow \hb\ component which is
forced to have the same velocity width and shift as those of \oiii, supporting
that \oiii\ is not blue-shifted and appropriate for defining the systematic
redshift of this object.


\begin{thebibliography}{}
  \bibitem[Adhikari et al.(2016)]{adhikari16}Adhikari, T. P., R\'o\.za\'nska,
    A., Czerny, B., et al. 2016, \apj, 831, 68
  \bibitem[Antonucci(1993)]{antonucci93}Antonucci, R. 1993, \araa, 31, 473
  \bibitem[Arnaud(1996)]{arnaud96}Arnaud, K. A. 1996,
    Astronomical Data Analysis Software and Systems V, 17
  \bibitem[Barth et al.(2015)]{barth15}Barth, A. J., Bennert, V. N.,
    Canalizo, G., et al. 2015, \apjs, 217, 26
  \bibitem[Barth et al.(2013)]{barth13}Barth, A. J., Pancoast, A., Bennert,
    V. N., et al. 2013, \apj, 769, 128
  \bibitem[Baskin et al.(2014)]{baskin14}Baskin, A., Laor, A., \& Stern, J.
    2014, \mnras, 438, 604
  \bibitem[Bechtold et al.(2002)]{bechtold02}Bechtold, J., Dobrzycki, A.,
    Wilden, B., et al. 2002, \apjs, 140, 143
  \bibitem[Bentz et al.(2013)]{bentz13}Bentz, M. C., Denney, K. D., Grier, C.
    J., et al. 2013, \apj, 767, 149
  \bibitem[Bentz \& Manne-Nicholas(2018)]{bentz18}Bentz, M. C., \&
    Manne-Nicholas, E. 2018, \apj, 864, 146
  \bibitem[Bentz et al.(2009)]{bentz09}Bentz, M. C., Walsh, J. L., Barth, A.
    J., et al. 2009, \apj, 705, 199
  \bibitem[Bian et al.(2010)]{bian10}Bian, W.-H., Huang, K., Hu, C., et al.
    2010, \apj, 718, 460
  \bibitem[Bianchi et al.(2019)]{bianch19}Bianchi, S., Antonucci, R., Capetti,
    A., et al. 2019, \mnras, 488, L1
  \bibitem[Blandford \& McKee(1982)]{blandford82}Blandford, R. D., \& McKee,
    C. F. 1982, \apj, 255, 419
  \bibitem[Bon et al.(2018)]{bon18}Bon, E., Marziani, P., Berton, M., et al.
    2018, Revisiting Narrow-line Seyfert 1 Galaxies and Their Place in the
    Universe. 9-13 April 2018. Padova Botanical Garden, 7
  \bibitem[Bon et al.(2012)]{bon12}Bon, E., Jovanovi\'c, P., Marziani, P., et
    al. 2012, \apj, 759, 118
  \bibitem[Bon et al.(2016)]{bon16}Bon, E., Zucker, S., Netzer, H., et al.
    2016, \apjs, 225, 29
  \bibitem[Boroson(2005)]{boroson05}Boroson, T. 2005, \aj, 130, 381
  \bibitem[Boroson \& Green(1992)]{boroson92}Boroson, T. A., \& Green, R. F.
    1992, \apjs, 80, 109
  \bibitem[Boroson \& Lauer(2009)]{boroson09}Boroson, T. A., \& Lauer, T. R.
    2009, \nat, 458, 53
  \bibitem[Brenneman \& Reynolds(2006)]{brenneman06}Brenneman, L. W., \&
    Reynolds, C. S. 2006, \apj, 652, 1028
  \bibitem[Brotherton(1996)]{brotherton96}Brotherton, M. S. 1996, \apjs, 102,
    1
  \bibitem[Brotherton et al.(1994)]{brotherton94}Brotherton, M. S., Wills, B.
    J., Francis, P. J., et al. 1994, \apj, 430, 495
  \bibitem[Cackett et al.(2007)]{cackett07} Cackett, E. M., Horne, K., \&
    Winkler, H. 2007, \mnras, 380, 669
  \bibitem[Cardelli et al.(1989)]{cardelli89}Cardelli, J. A., Clayton, G. C.
    \& Mathis, J. S. 1989, \apj, 345, 245
  \bibitem[Chelouche et al.(2019)]{chelouche19}Chelouche, D., Pozo Nu{\~n}ez,
    F., \& Kaspi, S. 2019, Nature Astronomy, 3, 251
  \bibitem[Corbin(1995)]{corbin95}Corbin, M. R. 1995, \apj, 447, 496
  \bibitem[Czerny \& Hryniewicz(2011)]{czerny11}Czerny, B., \& Hryniewicz, K.
    2011, \aap, 525, L8 
  \bibitem[Denney et al.(2009)]{denney09}Denney, K. D., Peterson, B. M.,
    Pogge, R. W., et al.  2009, \apj, 704, L80
  \bibitem[De Rosa et al.(2018)]{derosa18}De Rosa, G., Fausnaugh, M. M.,
    Grier, C. J., et al. 2018, \apj, 866, 133
  \bibitem[Du et al.(2018a)]{du18a}Du, P., Brotherton, M. S., Wang, K., et al.
    2018a, \apj, 869, 142
  \bibitem[Du et al.(2018b)]{du18b}Du, P., Zhang, Z.-X., Wang, K., et al.
    2018b, \apj, 856, 6
  \bibitem[Du et al.(2014)]{du14}Du, P., Hu, C., Lu, K.-X., et al. 2014, \apj,
    782, 45
  \bibitem[Du et al.(2015)]{du15}Du, P., Hu, C., Lu, K.-X., et al. 2015, \apj,
    806, 22
  \bibitem[Du et al.(2016)]{du16} Du, P., Lu, K.-X., Hu, C., et al. 2016,
    \apj, 820, 27
  \bibitem[Dumont \& Collin-Souffrin(1990)]{dumont90}Dumont, A. M. \&
    Collin-Souffrin, S. 1990, \aap, 229, 313
  \bibitem[Edelson et al.(2019)]{edelson19}Edelson, R., Gelbord, J., Cackett,
    E., et al. 2019, \apj, 870, 123
  \bibitem[Edelson et al.(2002)]{edelson02}Edelson, R., Turner, T. J., Pounds,
    K., et al. 2002, \apj, 568, 610
  \bibitem[Eracleous et al.(2012)]{eracleous12}Eracleous, M., Boroson, T. A.,
    Halpern, J. P., \& Liu, J. 2012, \apjs, 201, 23
  \bibitem[Eracleous \& Halpern(1994)]{eracleous94}Eracleous, M., \& Halpern,
    J. P. 1994, \apjs, 90, 1
  \bibitem[Fausnaugh et al.(2017)]{fausnaugh17}Fausnaugh, M. M., Grier, C. J.,
    Bentz, M. C., et al. 2017, \apj, 840, 97
  \bibitem[Ferland et al.(2009)]{ferland09}Ferland, G. J., Hu, C., Wang,
    J.-M., et al. 2009, \apjl, 707, L82
  \bibitem[Gaskell(2009)]{gaskell09}Gaskell, C. M. 2009, \nar, 53, 140
  \bibitem[Gaskell \& Peterson(1987)]{gaskell87}Gaskell, C. M., \& Peterson,
    B.  M. 1987, \apjs, 65, 1
  \bibitem[Gaskell \& Sparke(1986)]{gaskell86}Gaskell, C. M., \& Sparke, L. S.
    1986, \apj, 305, 175
  \bibitem[Goad \& Korista(2014)]{goad14}Goad, M. R., \& Korista, K. T. 2014,
    \mnras, 444, 43
  \bibitem[Grier et al.(2013a)]{grier13a}Grier, C. J., Martini, P., Watson, L.
    C., et al. 2013a, \apj, 773, 90
  \bibitem[Grier et al.(2017a)]{grier17a}Grier, C. J., Pancoast, A., Barth, A.
    J., et al. 2017, \apj, 849, 146
  \bibitem[Grier et al.(2008)]{grier08}Grier, C. J., Peterson, B. M., Bentz,
    M. C., et al. 2008, \apj, 688, 837
  \bibitem[Grier et al.(2013b)]{grier13b}Grier, C. J., Peterson, B. M., Horne,
    K., et al. 2013b, \apj, 764, 47
  \bibitem[Grier et al.(2017b)]{grier17b}Grier, C. J., Trump, J. R., Shen, Y.,
    et al. 2017, \apj, 851, 21
  \bibitem[Halpern et al.(1996)]{halpern96}Halpern, J. P., Eracleous, M.,
    Filippenko, A. V., et al. 1996, \apj, 464, 704
  \bibitem[Ho \& Kim(2014)]{ho14}Ho, L. C., \& Kim, M. 2014, \apj, 789, 17
  \bibitem[Ho et al.(2000)]{ho00}Ho, L. C., Rudnick, G., Rix, H.-W., et al.
    2000, \apj, 541, 120
  \bibitem[Horne(1994)]{horne94}Horne, K. 1994, in ASP Conf. Ser. 69,
    Reverberation Mapping of the Broadline Region in Active Galactic Nuclei,
    ed. P. M. Gondhalekar et al. (San Francisco, CA: ASP), 23
  \bibitem[Hu et al.(2015)]{hu15}Hu, C., Du, P., Lu, K.-X., et al. 2015,
    \apj, 804, 138
  \bibitem[Hu et al.(2020)]{hu20}Hu, C., Li, Y.-R., Du, P., et al. 2020,
    \apj, 890, 71
  \bibitem[Hu et al.(2008a)]{hu08a}Hu, C., Wang, J.-M., Ho, L. C., et al.
    2008a, \apj, 683, L115
  \bibitem[Hu et al.(2008b)]{hu08b}Hu, C., Wang, J.-M., Ho, L. C., et al.
    2008b, \apj, 687, 78 
  \bibitem[Hu et al.(2012)]{hu12}Hu, C., Wang, J.-M., Ho, L. C., et al. 2012,
    \apj, 760, 126
  \bibitem[Hu et al.(2016)]{hu16}Hu, C., Wang, J.-M., Ho, L. C., et al. 2016,
    \apj, 832, 197
  \bibitem[Kaspi et al.(2000)]{kaspi00}Kaspi, S., Smith, P. S., Netzer, H.,
    et al. 2000, \apj, 533, 631
  \bibitem[Koratkar \& Gaskell(1991)]{koratkar91}Koratkar, A. P., \& Gaskell,
    C. M. 1991, \apjs, 75, 719
  \bibitem[Kova\v{c}evi\'c et al.(2010)]{kovacevic10}Kova\v{c}evi\'c, J.,
    Popovi\'c, L. \v{C}., \& Dimitrijevi\'c, M. S.\ 2010, \apjs, 189, 15
  \bibitem[Korista \& Goad(2001)]{korista01}Korista, K. T. \& Goad, M. R.
    2001, \apj, 553, 695
  \bibitem[Korista \& Goad(2019)]{korista19}Korista, K. T. \& Goad, M. R.
    2019, \mnras, 489, 5284
  \bibitem[Lawther et al.(2018)]{lawther18}Lawther, D., Goad, M. R., Korista,
    K. T., et al. 2018, \mnras, 481, 533
  \bibitem[Le \& Woo(2019)]{le19}Le, H. A. N., \& Woo, J.-H.\ 2019, \apj, 887,
    236
  \bibitem[Li et al.(2018)]{li18}Li, Y.-R., Songsheng, Y.-Y., Qiu, J., et al.
    2018, \apj, 869, 137
  \bibitem[Li et al.(2016)]{li16} Li, Y.-R., Wang, J.-M., Ho, L. C., et al.
    2016, \apj, 822, 4
  \bibitem[Mangham et al.(2019)]{mangham19}Mangham, S. W., Knigge, C.,
    Williams, P., et al. 2019, \mnras, 488, 2780
  \bibitem[Maoz \& Netzer(1989)]{maoz89}Maoz, D., \& Netzer, H. 1989, \mnras,
    236, 21
  \bibitem[Marziani et al. (2009)]{marziani09}Marziani, P., Sulentic, J. W.,
    Stirpe, G. M., Zamfir, S., \& Calvani, M. 2009, \aap, 495, 83
  \bibitem[Murray et al.(1995)]{murry95}Murray, N., Chiang, J., Grossman, S.
    A., \& Voit, G. M. 1995, \apj, 451, 498
  \bibitem[Netzer(2020)]{netzer20}Netzer, H. 2020, \mnras, 494, 1611
  \bibitem[Netzer \& Marziani(2010)]{netzer10}Netzer, H., \& Marziani, P.
    2010, \apj, 724, 318
  \bibitem[O'Donnell(1994)]{odonnell94}O'Donnell, J. E. 1994, \apj, 422, 158
  \bibitem[Onken et al.(2004)]{onken04}Onken, C. A., Ferrarese, L., Merritt,
    D., et al. 2004, \apj, 615, 645
  \bibitem[Osterbrock \& Pogge(1985)]{osterbrock85}Osterbrock, D. E., \&
    Pogge, R. W. 1985, \apj, 297, 166
  \bibitem[Pancoast et al.(2011)]{pancoast11}Pancoast, A., Brewer, B. J., \&
    Treu, T. 2011, \apj, 730, 139
  \bibitem[Pancoast et al.(2014)]{pancoast14}Pancoast, A., Brewer, B. J.,
    Treu, T., et al. 2014, \mnras, 445, 3073
  \bibitem[Peterson(2014)]{peterson14}Peterson, B. M.  2014, \ssr, 183, 253
  \bibitem[Peterson et al.(2004)]{peterson04} Peterson, B. M., Ferrarese, L.,
    Gilbert, K. M., et al. 2004, \apj, 613, 682
  \bibitem[Peterson et al.(1998)]{peterson98}Peterson, B. M., Wanders, I.,
    Horne, K., et al. 1998, \pasp, 110, 660
  \bibitem[Popovi\'c et al.(2004)]{popovic04}Popovi\'c, L. \v{C}., Mediavilla,
    E., Bon, E., et al. 2004, \aap, 423, 909
  \bibitem[Popovi\'c et al.(2019)]{popovic19}Popovi\'c, L.  \v{C}.,
    Kova\v{c}evi\'c-Doj\v{c}inovi\'c, J., \& Mar\v{c}eta-Mandi\'c, S. 2019,
    \mnras, 484, 3180
  \bibitem[Rodr\'{i}guez-Pascual et
    al.(1997)]{rodriguez97}Rodr\'{i}guez-Pascual, P. M., Alloin, D., Clavel,
    J., et al. 1997, \apjs, 110, 9
  \bibitem[Rokaki et al.(1992)]{rokaki92}Rokaki, E., Boisson, C., \&
    Collin-Souffrin, S. 1992, \aap, 253, 57
  \bibitem[Schlafly \& Finkbeiner(2011)]{schlafly11}Schlafly, E. F., \&
    Finkbeiner, D. P. 2011, \apj, 737, 103
  \bibitem[Schmidt \& Green(1983)]{schmidt83}Schmidt, M., \& Green, R. F.
    1983, \apj, 269, 352
  \bibitem[Shields et al.(2000)]{shields00}Shields, J. C., Rix, H.-W.,
    McIntosh, D. H., et al. 2000, \apjl, 534, L27
  \bibitem[Storchi-Bergmann et al.(2017)]{storchi-bergmann17}Storchi-Bergmann,
    T., Schimoia, J. S., Peterson, B. M., et al. 2017, \apj, 835, 236
  \bibitem[Strateva et al.(2003)]{strateva03}Strateva, I. V., Strauss, M. A.,
    Hao, L., et al. 2003, \apj, 126, 1720
  \bibitem[Sulentic, \& Marziani(1999)]{sulentic99}Sulentic, J. W., \&
    Marziani, P. 1999, \apjl, 518, L9
  \bibitem[Sulentic et al.(2012)]{sulentic12}Sulentic, J. W., Marziani, P.,
    Zamfir, S., et al. 2012, \apjl, 752, L7
  \bibitem[Sulentic et al.(2000)]{sulentic00}Sulentic, J. W., Marziani, P.,
    Zwitter, T., Dultzin-Hacyan, D., \& Calvani, M. 2000, \apj, 545, L15
  \bibitem[Veilleux \& Osterbrock(1987)]{veilleux87}Veilleux, S., \&
    Osterbrock, D. E. 1987, \apjs, 63, 295
  \bibitem[Wang et al.(2012)]{wang12}Wang, J.-M., Du, P., Baldwin, J. A., et
    al. 2012, \apj, 746, 137
  \bibitem[Wang et al.(2017)]{wang17}Wang, J.-M., Du, P., Brotherton, M. S.,
    et al. 2017, NatAs, 1, 775
  \bibitem[Wang et al.(2014)]{wang14}Wang, J.-M., Qiu, J., Du, P. \& Ho, L. C.
    2014, \apj, 797, 65
  \bibitem[Wang et al.(2018)]{wang18}Wang, J.-M., Songsheng, Y.-Y., Li, Y.-R.,
    et al. 2018, \apj, 862, 171
  \bibitem[Welsh(1999)]{welsh99}Welsh, W. F. 1999, \pasp, 111, 1347
  \bibitem[White \& Peterson(1994)]{white94}White, R. J., \& Peterson, B. M.
    1994, \pasp, 106, 879
  \bibitem[Winkler \& Paul (2017)]{winkler17}Winkler, H., \& Paul, B. 2017,
    arXiv:1708.02056
  \bibitem[Xiao et al.(2018a)]{xiao18a}Xiao, M., Du, P., Horne, K., et al.
    2018, \apj, 864, 109
  \bibitem[Xiao et al.(2018b)]{xiao18b}Xiao, M., Du, P., Lu, K.-K., et al.
    2018, \apj, 865, L8
  \bibitem[Zhang(2013)]{zhang13}Zhang, X.-G. 2013, \mnras, 434, 2664
  \bibitem[Zhou et al.(2019)]{zhou19}Zhou, H., Shi, X., Yuan, W., et al. 2019,
    \nat, 573, 83
  \bibitem[Zu et al.(2011)]{zu11}Zu, Y., Kochanek, C. S., \& Peterson, B. M.
    2011, \apj, 735, 80
\end{thebibliography}
\end{document}